\documentclass[11pt]{article}

\renewcommand{\theequation}{\thesection.\arabic{equation}}

\usepackage{graphicx}

\newcommand{\graphfile}[1]{\centerline{\includegraphics[width=0.86\textwidth]{#1}}}
\usepackage[numbers,sort&compress]{natbib}
\newcommand{\onlinecite}[1]{\cite{#1}}

\usepackage{amsmath}
\usepackage{amsthm}

\theoremstyle{definition}
\newtheorem*{maintheorem}{Theorem}
\newtheorem*{definition}{Definition}
\newtheorem{theorem}{Theorem}

\newtheorem{lemma}{Lemma}

\newcommand{\eq}[1]{\eqref{#1}}
\newcommand{\Eq}[1]{Eq.~\eq{#1}}
\newcommand{\Eqs}[1]{Eqs.~\eq{#1}}

\newcommand{\keywords}[1]{\par\noindent{\footnotesize\bf KEY WORDS:} #1}
\newcommand{\pacs}[1]{\par\noindent{\footnotesize\bf PACS CODES:} #1}

\newcommand{\average}[1]{\langle#1\rangle}
\newcommand{\baverage}[1]{\Big\langle#1\Big\rangle}
\newcommand{\cumulant}[1]{\langle\!\langle#1\rangle\!\rangle}
\newcommand{\bcumulant}[1]{\Big\langle\!\!\Big\langle#1\Big\rangle\!\!\Big\rangle}
\newcommand{\dd}[3][]{\frac{\partial^{#1} #2}{\partial #3^{#1}}}
\newcommand{\ddh}[3][]{\frac{d^{#1} #2}{d #3^{#1}}}
\newcommand{\order}[1]{\mathcal O(#1)}

\newcommand{\Theorem}[1]{Theorem~\ref{#1}}
\newcommand{\Lemma}[1]{Lemma~\ref{#1}}
\newcommand{\Sec}[1]{Part~\ref{#1}}
\newcommand{\Secp}[1]{Part~\ref{#1} of the paper}
\newcommand{\Subsec}[1]{Sec.~\ref{#1}}
\newcommand{\textorder}[3][order]{the $#3$-th #1 in $#2$}
\newcommand{\thiscumulant}
{\cumulant{V_1^{[p_1]};\ldots;V_{\mathcal N}^{[p_{\mathcal N}]}
;P_1^{[n_1]};\ldots;P_Q^{[n_Q]}} }
\newcommand{\thisbcumulant}
{\cumulant{V_1^{[p_1]};\ldots;V_{\mathcal N}^{[p_{\mathcal N}]}
;P_1^{[n_1]};\ldots;P_Q^{[n_Q]}} }

\begin{document}

\title{Theorem on the Distribution of Short-Time Particle
Displacements with Physical Applications}

\author{R. van Zon and E. G. D. Cohen\\
\emph{The Rockefeller University, New York, NY 10021, USA}} 

\date{11 October 2005}

\maketitle

\begin{abstract}\noindent
The distribution of the initial short-time displacements of particles
is considered for a class of classical systems under rather general
conditions on the dynamics and with Gaussian initial velocity
distributions, while the positions could have an arbitrary
distribution.  This class of systems contains canonical equilibrium of
a Hamiltonian system as a special case.  We prove that for this class
of systems the $n$th order cumulants of the initial short-time
displacements behave as the $2n$-th power of time for all $n>2$,
rather than exhibiting an $n$th power scaling.  This has direct
applications to the initial short-time behavior of the Van Hove
self-correlation function, to its non-equilibrium generalizations the
Green's functions for mass transport, and to the non-Gaussian
parameters used in supercooled liquids and glasses.



\end{abstract}

\section*{On the structure of this paper}
\label{structure}

This paper concerns a universal property of correlations of the
initial short-time behavior of the displacements of particles for a
class of ensembles of classical systems, both in and out of
equilibrium (the latter one being somewhat restricted). Among other
things, these correlations (expressed in terms of the so-called
cumulants) have applications to neutron
scattering\cite{VanHove54,Schofield61,Rahmanetal62,Rahman64,NijboerRahman66,LevesqueVerlet70,Sears72,BoonYip80,HansenMcDonald86,Squires,Arbeetal02,Arbeetal03},
to the description of a restricted class of non-equilibrium systems on
small time and length
scales\cite{Kincaid95,KincaidCohen02a,KincaidCohen02b,CohenKincaid02,VanZonCohen05b},
and to heterogeneous dynamics in supercooled liquids and
glasses\cite{OdagakiHiwatari91,KobAndersen95,HurleyHarrowell96,Kobetal97,ZangiRice04}.

The property can be formulated as a mathematical theorem, which may
have more applications than are considered in this paper. Our interest
is however mainly in its physical applications.  Since the general
theorem and its proof are best formulated in a rather abstract way
which is not necessary for the currently known physical applications,
the paper is split up into a physical part (\Sec{physics}) containing
the physical formulation of the Theorem and its applications, and a
mathematical part (\Sec{mathematics}) containing the general
mathematical formulation of the Theorem and its proof.

These two parts are further subdivided into sections. In
\Subsec{introduction} the physical motivation for studying particle
displacements is given.  In \Subsec{systems} we introduce the physical
systems that will be considered. Section~\ref{moments and cumulants}
gives the necessary definitions to be able to treat the cumulants of
the particle displacements.  We state the Theorem in physical terms in
\Subsec{thephysicaltheorem} and discuss its physical applications in
\Subsec{applications}.

As for the mathematical \Secp{mathematics}, in \Subsec{general
definition}, we give the general definition of cumulants of any number
of random variables and discuss some of their properties, while in
\Subsec{thetheorem} we present the Theorem in its full mathematical
generality.  In \Subsec{proof} we prove the Theorem. In the proof we
need an auxiliary theorem concerning Gaussian distributed variables
whose proof is postponed to the Appendix. Some coefficients occurring
in the Theorem are worked out in \Subsec{coefficients}.

We end with  conclusions, which are followed by the Appendix.

\section{Physical Part}
\label{physics}

\subsection{Introduction}
\label{introduction}

Our motivation to consider the displacements of particles originates
from trying to describe the behavior of non-equilibrium systems on all
time scales using the so-called Green's function theory, introduced by
J.~M.\ Kincaid\cite{Kincaid95}. This theory aims to describe,
among other things, the time evolution of the number densities,
momentum density and energy density, by expressing them in terms of
Green's functions. It has so far been successfully applied to
self-diffusion\cite{Kincaid95} and to heat
transport\cite{KincaidCohen02a,KincaidCohen02b,CohenKincaid02} while a
study of mass transport in binary (isotopic)
mixtures\cite{VanZonCohen05b} is in progress. The main advantage of
the Green's functions over hydrodynamics is that one expects that they
can describe the system on more than just long time and
length scales, in particular also on time scales of the order of
picoseconds and on length scales of the order of nanometers. The
connection between the picosecond and nanometer scale can be
understood by realizing that with typical velocities of 500 m/s, a
particle in a fluid at room temperature moves about 0.5 nm in 1 ps.
Hence studying the picosecond and sub-picosecond time scales could
also be relevant for nanotechnology.

For mass transport in multi-component
fluids\cite{Kincaid95,VanZonCohen05b}, the Green's functions
$G_\lambda(\mathbf r,\mathbf r',t)$ have the physical interpretation
of being the probability that a single particle of component $\lambda$
is at a position $\mathbf r$ at time $t$ given that it was at a
position $\mathbf r'$ at time zero. At time zero the system is not in
equilibrium, in fact, a class of far-from-equilibrium
situations has been studied in this
context\cite{KincaidCohen02a,KincaidCohen02b,CohenKincaid02,VanZonCohen05b}.
Given this interpretation, it is clear that the displacements of
single particles are the central quantities in the Green's function
theory.

Apart from the non-equilibrium aspect, the above interpretation of the
Green's functions is the same as that of the classical equilibrium Van
Hove self-correlation function $G_s(\mathbf r-\mathbf
r',t)$\cite{VanHove54,BoonYip80,HansenMcDonald86}. In the literature
on the classical Van Hove self-correlation function in the context of
neutron scattering on an equilibrium
fluid\cite{VanHove54,BoonYip80,Squires}, the so-called cumulants of
the particle displacements (defined in \Subsec{moments and cumulants}
below) have been studied by Schofield\cite{Schofield61} and
Sears\cite{Sears72}, among others.  The relevance of the cumulants for
the Van Hove self-correlation function can be seen from its Fourier
transform, the incoherent intermediate scattering
function\cite{VanHove54,BoonYip80,Squires,HansenMcDonald86}, which can
be measured and is defined as
\begin{equation}
  F_s(k,t) = \baverage{e^{\mathrm{i} k
\hat{\mathbf{k}}\cdot[\mathbf{r}_1(t)-\mathbf{r}_1(0)]} }_\text{eq} =
\baverage{e^{\mathrm{i} k \Delta x_1(t)}}_\text{eq},
\label{Fsdef}
\end{equation}
where $\mathbf k=k\,\hat{\mathbf k}$ is a wave vector with length $k$
and direction $\hat{\mathbf{k}}$ (a unit
vector), $\mathbf r_1(t)$ the position of a (single) particle at
time $t$ and $\average{}_\text{eq}$ an equilibrium average.
In the last equality we have chosen
$\hat{\mathbf{k}}=\hat{\mathbf{x}}$ (in an isotropic equilibrium fluid
the result is independent of the direction $\hat{\mathbf k}$) and
defined $\Delta x_1(t)$ as the displacement of the single particle in
the $x$ direction: $\Delta x_1(t)=\hat{\mathbf x}\cdot[\mathbf r_1(t)-
\mathbf r_1(0)]$.  In probability theory\cite{Cramer46,VanKampen}
$\log\average{\exp[\mathrm{i} k A]}$ is the cumulant generating
function for the random variable $A$, so we see from \Eq{Fsdef}
that $\log F_s(k,t)$ is the cumulant generating function of $\Delta
x_1(t)$. Note that $\Delta x_1(t)$ is a random variable here as it depends
on an initial phase point drawn from a probability distribution (here
the equilibrium distribution). The cumulant generating function is,
by definition, equal to $\sum_{n=1}^\infty\kappa_n (\mathrm{i}
k)^n/n!$ where $\kappa_n$ is the $n$th
cumulant\cite{Cramer46,VanKampen}. The relation of the incoherent
intermediate scattering function and the cumulants of the displacement
is thus expressed
by\cite{Rahmanetal62,NijboerRahman66,Sears72,BoonYip80,DeSchepperetal81}
\begin{equation}
  F_s(k,t) = 
\exp \sum_{n=1}^\infty \frac{\kappa_n}{n!}(\mathrm{i} k)^n.
\label{Fsrelation}
\end{equation}
This connection with the incoherent scattering function (and thus with
neutron
scattering\cite{VanHove54,Schofield61,Rahmanetal62,Rahman64,NijboerRahman66,LevesqueVerlet70,Sears72,BoonYip80,HansenMcDonald86,Squires,Arbeetal02,Arbeetal03})
explains the early interest in the cumulants of displacements from a
physical perspective.  We note that \Eq{Fsrelation} also shows
that the cumulants could be obtained experimentally by measuring
$F_s(k,t)$ via small $k$ (``small angle'') incoherent neutron
scattering and fitting the logarithm of the result to a power series
in $k$.  In \Subsec{applications} we will discuss also more recent
physical applications of the cumulants such as the non-equilibrium
Green's functions and non-Gaussian parameters used in the theory of
supercooled liquids and glasses.

In the context of incoherent neutron scattering, it has been noted
by Schofield\cite{Schofield61} and Sears\cite{Sears72} that the
cumulants $\kappa_n$ of the displacement of a particle in a time $t$
in a fluid in equilibrium with a smooth interparticle potential behave
for small $t$ as $\kappa_2 = \order{t^2}$, $\kappa_4 = \order{t^8}$,
$\kappa_6 = \order{t^{12}}$, while all odd cumulants~vanish.  These
results suggested for equilibrium systems with smooth potentials a
behavior as \textorder[power]{t}{2n} for $\kappa_n$ when $n>2$ in
general, but to the best of our knowledge no proof of this property is
available at present.

We note that an $\order{t^{2n}}$ behavior would be in stark contrast
to results obtained for hard disk and hard sphere fluids in
equilibrium\cite{Sears72,DeSchepperetal81}. Sears\cite{Sears72}
considered results up to \textorder{t}{8} and \textorder{t}{12} for
$\kappa_4$ and $\kappa_6$, respectively, for smooth potentials and he
found by using a limit in which the smooth potential reduces to a hard
core potential that for hard spheres $\kappa_4=\order{|t|^5}$ and
$\kappa_6=\order{|t|^7}$, while the odd cumulants were still zero.  An
alternative approach was followed by De Schepper et
al.\cite{DeSchepperetal81}\ consisting of directly evaluating the Van
Hove self-correlation function for short times for hard spheres based
on pseudo-Liouville operators (which replace the usual ones for smooth
potentials). De Schepper et al.\cite{DeSchepperetal81}\ obtained
$\kappa_n = \order{|t|^{n+1}}$ for all even $n>2$, with corrections of
$\order{|t|^{n+2}}$.

The $\order{t^{2n}}$ result for smooth potentials is the more
remarkable in that a naive estimate of the short time behavior of
$\kappa_n$ based on its connection with the moments would predict a
behavior as \textorder{t}{n}.  Hence \emph{all} terms from
$\order{t^n}$ up to $\order{t^{2n-1}$},\footnote{Everywhere in this
paper ``up to $\order{t^\alpha}$'' means ``up to and including
$\order{t^\alpha}$''.}  should \emph{vanish}.  The question we address
here is whether this is indeed general for smooth potentials.

In fact in this paper we shall show that under quite general
conditions, the main one being that in the initial ensemble, the
velocities have a multi-variate Gaussian distribution\cite{VanKampen}
and are statistically independent of the positions, one can prove a
very general theorem to be presented in the next part. This theorem
implies, among other things, that for a many particle system in which
the (interparticle as well as external) forces are independent of the
velocities, the $n$th order cumulants of the displacement are indeed
$\order{t^{2n}}$ for $n>2$ and $\order{t^{n}}$ for $n\leq 2$. This
confirms the above stated expectation, but extends it to a restricted
class of non-equilibrium situations as well, as we will show.

\subsection{Class of physical systems}
\label{systems}

In this part we will restrict ourselves to the following class of
systems (the Theorem in \Secp{mathematics} concerns a more
general class).  Consider $N$ point particles in three
dimensions\footnote{Other dimensionalities work just as well.} whose
positions and velocities are denoted by the three-dimensional vectors
$\mathbf r_i$ and $\mathbf v_i$, respectively, and whose masses are
$m_i$, where $i=1\ldots
N$. The time evolution of the particles is governed by
\begin{eqnarray}
  \dot{\mathbf r}_i &=& \mathbf v_i
\label{eqmotion1}
\\
  \dot{\mathbf v}_i &=& \mathbf F_i(\mathbf r^N,t)/m_i,
\label{eqmotion2}
\end{eqnarray}
where $\mathbf F_i$ is the force acting on particle $i$, which is
supposed to be a smooth function of the time $t$ and of $\mathbf r^N$,
which is the collection of all positions $\mathbf r_i$. Likewise,
$\mathbf v^N$ will denote the collection of all velocities $\mathbf
v_i$.

We consider an ensemble of such systems. In the ensemble, the initial
probability distribution $\mathcal P(\mathbf r^N,\mathbf v^N)$ of the positions
and the velocities of the particles is such that the velocities each
have a Gaussian distribution and are statistically independent of the
positions, i.e.,\footnote{The expression in \Eq{physicaldistribution}
is not the most general Gaussian distribution for $\mathbf v^N$. While
the Theorem in \Secp{mathematics} allows any multivariate Gaussian
distribution of the velocities, we have in fact not found any physical
applications for that yet.}
\begin{equation}
  \mathcal P(\mathbf r^N,\mathbf v^N)=
  f(\mathbf r^N)
 \prod_{i=1}^N 
\left[
\left(\frac{\beta_im_i}{2\pi}\right)^{3/2}
\exp\Big(-\frac12{\beta_im_i}|\mathbf v_i-\mathbf
  u_i|^2\Big)
\right].
\label{physicaldistribution}
\end{equation}
where $\mathbf u_i$ is the average of the initial velocity $\mathbf
v_i$ and $\beta_{i}$ are positive ``inverse temperature''-like
variables. In \Eq{physicaldistribution}, the $f(\mathbf r^N)$ denotes
a general probability distribution function of the initial positions
of the particles.  While the distribution function in
\Eqs{physicaldistribution} can describe both equilibrium and non-equilibrium
situations, it always shares with the equilibrium distribution the Gaussian
dependence on the velocities, which is in fact crucial for the Theorem
below to hold.

We stress that the forces $\mathbf F_i$ in \Eq{eqmotion2} are
independent of the velocities, but may depend on the positions
$\mathbf r^N$ and on the time $t$ in any way as long as they are
smooth. An example of smooth forces would be infinitely differentiable
forces $\mathbf F_i(\mathbf r^N,t)$, but also Lennard-Jones-like forces are
allowed provided the distribution of the positions $f(\mathbf r^N)$
assigns a vanishing probability to the particles to be at zero
distance of one another, which is the singular point of the
Lennard-Jones-like potential at which it is not smooth.

Canonical equilibrium for a single or multi-component fluid is just a
special case of these systems. In that case, one has $\mathbf u_i=0$,
$\beta_{i}=\beta$, $\mathbf F_i=-\partial U/\partial\mathbf r_i$ and
$f(\mathbf r^N)\propto\exp[-\beta U(\mathbf r^N)]$, where $U(\mathbf
r^\mathcal N)$ is the potential energy function of the system.

The probability distribution functions of the form
\Eq{physicaldistribution} in which each particle has its own mean
velocity $\mathbf u_i$ and ``inverse temperature'' $\beta_i$, may seem
at first of a mathematical generality which has little physical
relevance. Note, however, that this is a convenient way to describe
mixtures of any arbitrary number of components. In such a mixture, the
mean velocities and/or temperatures of the different components could be
selected physically e.g.\ by having two vessels with different
substances at different temperatures with a divider which is opened at
$t=0$, or by means of a laser (or perhaps even a neutron beam) applied
at $t=0$, tuned to a resonance of one of the components only, which
would give the particles of that component a nonzero average momentum
as well as a different initial ``temperature'' due to the
recoil energy\cite{Squires}. To what extent such techniques could
ensure the Gaussianity of the velocity distribution is a technical
matter that we will not go into here.

\subsection{Definition of moments and cumulants}
\label{moments and cumulants}

In this paper, when we speak of a random variable, we mean a
function of the positions $\mathbf r^N$ and the velocities $\mathbf
v^N$ and possibly the time $t$. All physical quantities are variables
of this kind.  For such variables $A(\mathbf r^N,\mathbf v^N,t)$, the
average with respect to the distribution function $\mathcal P(\mathbf
r^N,\mathbf v^N)$ will be denoted by
\begin{equation}
\average{A} = \int\! d\mathbf r^Nd\mathbf v^N\, 
\mathcal P(\mathbf r^N,\mathbf v^N) A(\mathbf r^N,\mathbf v^N,t).
\label{average}
\end{equation}
Throughout the paper, $A$'s will denote general variables.

We define $\Delta\mathbf r_i(t)=(\Delta x_i(t), \Delta y_i(t), \Delta
z_i(t))$ as the displacement of particle~$i$ in time $t$, where
$i=1\ldots N$ and the time dependence is suppressed for brevity.  The
$n$th moment\footnote{Although the moments $\mu_n$ and cumulants
$\kappa_n$ are really the ``moments and cumulants of the probability
distribution function of $\Delta x_i(t)$,'' we will refer to them simply
as the ``moments and cumulants of $\Delta x_i(t)$.''} of $\Delta x_i(t)$ is
the average of its $n$th power, and is denoted as
\begin{equation}
  \mu_n \equiv \average{\Delta x^n_i(t)},
\label{moments}
\end{equation}
where the dependence on $i$ and $t$ on the left-hand side (lhs) has
been suppressed.  In the following we will also suppress the $t$
dependence in $\Delta x_i$.  The cumulants of $\Delta
x_i$,\footnotemark[\value{footnote}] denoted by $\kappa_n$, are equal
to the moments $\mu_n$ with certain factorizations of them subtracted
and thus sensitive to the correlations of $\Delta x_i$. Their precise
definition is via the cumulant generating
function\cite{Cramer46,VanKampen}
\begin{equation}
  \Phi(k) \equiv \log\baverage{\exp[\mathrm{i} k \Delta x_i]}
  = 
  \sum_{n=1}^\infty  \frac{\kappa_n}{n!}(\mathrm{i} k)^n.
\label{cumulantgenerator}
\end{equation}
from which the $\kappa_n$ follow as
\begin{equation}
\kappa_n \equiv \dd[n]{\Phi}{(\mathrm{i} k)}\bigg|_{k=0}\:.
\label{kappan}
\end{equation}
An alternative notation for the cumulants, which is more analogous to
\Eq{moments} and which is especially convenient in the case of several
variables, is\cite{VanKampen}
\begin{equation}
 \cumulant{\Delta x_i^{[n]}}\equiv \kappa_n.
\label{cumulantnot}
\end{equation}
We stress here that the superscript $[n]$ is not a power. Furthermore,
we will follow the convention that the superscript will not be denoted
if $n=1$, i.e., $ \cumulant{\Delta x_i} \equiv \cumulant{\Delta
x_i^{[1]}}$ (which is also equal to $\kappa_1 = \average{\Delta
x_i}$).

The cumulant generating function in \Eq{cumulantgenerator} can be
expressed in terms of the moments $\mu_n$ since
$\Phi(k)=\log\average{\exp[\mathrm{i} k \Delta
x_i]}=\log[1+\sum_{n=1}^\infty (\mathrm i k)^n\mu_n/n!]$. The
relations between the cumulants and moments can then be found by
Taylor expanding the logarithm around $1$ and using a multinomial
expansion for the powers of the second term inside the logarithm. This
gives, somewhat formally, 
\begin{equation}
  \kappa_n \:=\: -\,n! 
\mathop{\sum_{\{p_\ell\geq 0\}}}_{\sum_{\ell=1}^\infty \ell p_\ell = n}
\Big(\sum_{\ell=1}^\infty p_\ell-1\Big)!
\prod_{\ell=1}^\infty \frac{\big({-\mu_\ell}/{\ell!}\big)^{p_\ell}
}{p_\ell!}
\label{kappaintermsofmu}
\end{equation}
For instance, for the first few $\kappa_n$, \Eq{kappaintermsofmu}
becomes $\kappa_1=\mu_1$, $\kappa_2=\mu_2-\mu_1^2$ and
$\kappa_3=\mu_3-3\mu_1\mu_2+2\mu_1^3$\cite{Cramer46,VanKampen}.

In general, $\kappa_n=\mu_n\pm$ factored terms, where the factored
terms contain all ways of partitioning the moment $\mu_n$ into a
product of lower moments $\mu_\ell$ such that all $\ell$ values
(taking into account their ``frequencies of occurrence'' $p_\ell$) add
up to $n$.

Apart from the cumulants $\kappa_n$ of the single variable $\Delta
x_i$, we will also need the general definition of cumulants which
applies to the displacement of the same particle in different spatial
directions as well as the displacements of different particles.  Like
for the cumulants of a single particle's displacement, the expressions
for the general cumulants are moments with factored forms subtracted,
e.g.
\begin{eqnarray}
  \cumulant{\Delta x_1;\Delta x_2} 
&=& \average{\Delta x_1\Delta x_2}
   -\average{\Delta x_1}\average{\Delta x_2}
\label{cumulantex1}
\\
  \cumulant{\Delta x_1;\Delta y_1;\Delta z_1} 
&=&
  \average{\Delta x_1\Delta y_1\Delta z_1} 
  -\average{\Delta x_1}\average{\Delta y_1\Delta z_1} 
  -\average{\Delta x_1\Delta y_1}\average{\Delta z_1} 
\nonumber\\&&
  -\average{\Delta x_1\Delta z_1}\average{\Delta y_1}
  +2\average{\Delta x_1}\average{\Delta y_1}\average{\Delta z_1}
\label{cumulantex2}
\\
  \cumulant{\Delta x_1^{[2]};\Delta y_1^{[2]}} &=&  
\average{\Delta x_1^2\Delta y_1^2} 
-
\average{\Delta x_1^2}\average{\Delta y_1^2} 
-2 \average{\Delta x_1\Delta y_1}^2  
\nonumber\\
&&
-2 \average{\Delta x_1}\average{\Delta x_1\Delta y_1^2} 
-2 \average{\Delta x_1^2\Delta y_1}\average{\Delta y_1}
\nonumber\\&&
+8\average{\Delta x_1}\average{\Delta x_1\Delta y_1}\average{\Delta
  y_1}
-6\average{\Delta x_1}^2\average{\Delta y_1}^2.
\label{cumulantex3}
\end{eqnarray}
Here, the semi-colons are inserted  to avoid ambiguity  and to make it
explicit that the expressions inside the double angular brackets are
not to be multiplied. This is different from the notation for the
multivariate cumulants used by Van Kampen~\cite{VanKampen} who
denotes the above cumulants as $\cumulant{\Delta x_1\Delta x_2}$,
$\cumulant{\Delta x_1\Delta y_1\Delta z_1}$ and $\cumulant{\Delta
x_1^2\Delta y_1^2}$, respectively, which would suggest that the
expressions inside the double brackets are products of powers, which
they are not. 

The general definition of the
multivariate cumulants will be given in \Eq{cumintermsofmom} in
\Subsec{general definition} (where we will also show how the various
prefactors are determined) and can also be found in
refs.~\onlinecite{Cramer46} and \onlinecite{VanKampen}.

\pagebreak[3]
\subsection{The Theorem in physical terms}

\label{thephysicaltheorem}

For the class of physical systems described above, the general and
somewhat formal Theorem in \Secp{mathematics} states that, in a less
abstract physical notation
\begin{multline}
\cumulant{\Delta x_1^{[n_{1x}]};
\Delta y_1^{[n_{1y}]};\Delta z_1^{[n_{1z}]}
\Delta x_2^{[n_{2x}]};
\Delta y_2^{[n_{2y}]};\Delta z_2^{[n_{2z}]};\cdots
}
\\
\:=\:  \begin{cases}
               c_{\{n_{i\eta}\}} t^{n}+ \order{t^{n+1}} &  \text{if }\: n\leq2
		\\
               c_{\{n_{i\eta}\}} t^{2n}+ {\order{t^{2n+1}}} &   \text{if }\: n>2,
             \end{cases}
\label{phystheorem}
\end{multline}
where $n_{i\eta}$ are non-negative numbers (some may be zero),
$i=1\ldots N$, $\eta=x,y$ or $z$ and
\begin{equation}
  n = \sum_{i=1}^N\sum_{\eta=x,y,z} n_{i\eta}
\end{equation}
is the \emph{order of the cumulant} on the lhs\ of \Eq{phystheorem}.

Note that the cumulants $\kappa_n$ that occur in the
expansion of the Van Hove self-correlation function are special cases
of these $n$th order cumulants, i.e., we can write
\begin{equation}
\kappa_n
= \cumulant{\Delta x_i^{[n]}}
= \begin{cases}
\,c_nt^n\,+\order{t^{n+1}} &\quad \text{for $n\leq2$}\\
c_nt^{2n}+\order{t^{2n+1}} &\quad \text{for $n> 2$.}
\end{cases}
\label{Theorem0}
\end{equation}
Section~\ref{coefficients} contains the expressions for the $c_n$
and the $c_{\{n_i\}}$.  

We stress once more that on the basis of the connection between the
cumulants and the moments \eq{kappaintermsofmu}, only an $\order{t^n}$
scaling of the $n$th order cumulants is to be expected, so that this
is a nontrivial theorem.  The result in \Eq{Theorem0} generalizes the
results of Schofield\cite{Schofield61} and Sears\cite{Sears72} who
found that for an equilibrium liquid, $\kappa_2=\order{t^2}$ while
$\kappa_4=\order{t^8}$ and $\kappa_6=\order{t^{12}}$. The results in
\Eqs{phystheorem} and \eq{Theorem0} hold, however, also out of
equilibrium as long as the initial distribution of the particle
velocities is Gaussian and statistically independent of the particles'
positions.

\subsection{Applications}
\label{applications}

We will now discuss a number of physical applications of the Theorem in
\Eqs{phystheorem}--\eq{Theorem0} just
presented.

\subsubsection{The equilibrium Van Hove self-correlation function}

a) In incoherent neutron scattering on an equilibrium fluid, one
essentially measures the equilibrium Van Hove self-correlation
function $G_s(r,t)$ (with $r=|\mathbf r-\mathbf r'|$), which is the
Fourier inverse of the incoherent scattering function
$F_s(k,t)$\cite{VanHove54,BoonYip80,HansenMcDonald86,Squires,Arbeetal02,Arbeetal03}.
If the length of the wave vector $k$ in $F_s(k,t)$ is small, then
according to \Eq{Fsrelation} $F_s(k,t)\approx\exp[-\kappa_2k^2/2]$,
i.e., nearly Gaussian, and so its inverse Fourier transform $G_s(r,t)$
is also approximately Gaussian. Corrections to this Gaussian behavior
can be found by resumming $F_s(k,t)=\exp\sum_n \kappa_n(\mathrm i
k)^n/n!$ in \Eq{Fsrelation} to the form
\begin{equation}
F_s(k,t)=e^{-\frac12\kappa_2 k^2}\Bigg[1+\mathop{\sum_{n=4}}_{n\text{ even}}^\infty b_n
 k^n\Bigg]
\end{equation}
using that in equilibrium odd cumulants are zero. For $n\geq 4$, the
coefficients $b_{n}$ are given by\cite{VanZonCohen05b}
\begin{equation}
  b_{n} \:=\: 
\mathop{
\sum
_{\{p_\ell\geq 0;\: \ell\text{ even}\}}
}_{\sum_{\ell=4,\ell\text{ even}}^\infty \ell p_\ell = n}
\mathop{\prod_{\ell=4}}_{\ell\text{ even}}^\infty\:\:
\left[
\frac{1}{p_\ell!}\left(\frac{\kappa_{\ell}}{\ell!}\right)^{p_\ell}
\right]
.
\label{equilbnintermsofkappan}
\end{equation}
For example, $b_4=\kappa_4/4!$, $b_6=\kappa_6/6!$,
$b_8=(\kappa_8+35\kappa_4^2)/8!$.  Fourier inverting the resummed form
of $F_s(k,t)$ leads to
\begin{equation}
  G_s(r,t) = \frac{e^{-w^2}}{\sqrt{2\pi\kappa_2}}\left[
1+
\mathop{\sum_{n=4}^\infty}_{n\text{ even}}
\frac{b_{n}}{(2\kappa_2)^{n/2}}H_n(w)\right].
\label{equilGs}
\end{equation}
Here $H_n$ is the $n$th Hermite polynomial and the dimensionless
$w\equiv r/\sqrt{2\kappa_2}$.  We note that the series in \Eq{equilGs}
appears to have a fairly rapid convergence\cite{LevesqueVerlet70}.
Taking just the first few terms would give
\begin{equation}
  G_s(r,t) = \frac{e^{-w^2}}{\sqrt{2\pi\kappa_2}}\left[
1+\frac{\kappa_4}{4!4\kappa_2^2}H_4(w)+\frac{\kappa_6}{6!8\kappa_2^3}H_6(w)+
\ldots
\right].
\label{afewgoodterms}
\end{equation}

The Theorem in \Eq{Theorem0} provides a justification for the
expansion in \Eq{equilGs} of the self part of the Van Hove function
$G_s$ for short times $t$ in the following way.  As
\Eqs{equilbnintermsofkappan} and \eq{equilGs} show, the cumulants
$\kappa_{n\geq4}$ give, via the $b_n$, corrections to a Gaussian
behavior of $G_s(r,t)$.  The Gaussian factor $e^{-w^2}$ suggests that
typical values of $w$ are $\order{1}$ in \Eq{equilGs}, so also
$H_n(w)=\order{1}$. Its prefactor in \Eq{equilGs} is, however,
$t$-dependent through $b_n/(2\kappa_2)^{n/2}$. Given the relation
between $b_n$ and $\kappa_n$ in \Eq{equilbnintermsofkappan} it is easy
to see that they scale similarly, i.e., if the conditions of the
Theorem are satisfied so that $\kappa_{n>2}=\order{t^{2n}}$ then also
$b_{n>2}=\order{t^{2n}}$. Since $\kappa_2=\order{t^2}$, we obtain
\begin{equation}
    \frac{b_n}{(2\kappa_2)^{n/2}}  = \order{t^n}.
\end{equation}
\emph{This means that the series in \Eq{equilGs} is well-behaved for
small times $t$, in that each next term is smaller than the previous
one, and that by truncating the series\footnote{When truncating, 
terms that are kept have to include not just the leading
order in~$t$, but at least all terms  up to $\order{t^k}$ if terms up to $H_k$
are kept. In numerical approaches it is possible to retain the full
terms.} one obtains for small $t$
approximations which can be systematically improved by taking more
terms into account.}  Note that in contrast if $\kappa_n$ had been
$\order{t^n}$, each term in the series in \Eq{equilGs} would have
been of the same order.

b) An expansion of a similar form as \Eq{equilGs} was found by
Rahman\cite{Rahman64} for $G_s(r,t)$, and by Nijboer and
Rahman\cite{NijboerRahman66} for $F_s(k,t)$. Their expressions are in
terms of the so-called \emph{non-Gaussian parameters}
$\alpha_n$. These non-Gaussian parameters have recently also found 
applications in the context of supercooled liquids and glasses, where
they have been proposed as a kind of order parameter for the glass
transition\cite{OdagakiHiwatari91,Arbeetal02,Arbeetal03} and as
measures of ``dynamical heterogeneities'' in supercooled liquids and
glasses\cite{KobAndersen95,HurleyHarrowell96,Kobetal97,ZangiRice04}.
We note that while supercooled liquids and glasses are not in
true equilibrium (for that would be the solid phase), they do have a
Gaussian velocity distribution `inherited' from the fluid phase.

Given the present interest in these non-Gaussian parameters $\alpha_n$, we
will now compare them with the cumulants $\kappa_n$.  The non-Gaussian
parameters $\alpha_n$ are defined in terms of the distance $\Delta R$
$=$ $\sqrt{\Delta x^2(t)+\Delta y^2(t)+\Delta z^2(t)}$ traveled by any
particle in time $t$ in a three dimensional fluid, as\cite{Rahman64}
\begin{eqnarray}
  \alpha_n&\equiv&
\frac{\average{\Delta R^{2n}}}{\average{\Delta R^2}^n(2n+1)!!/3^n}-1.
\label{alphandef}
\end{eqnarray}
For isotropic fluids,
$\average{\Delta R^{2n}}=(2n+1)\average{\Delta x^{2n}(t)}=(2n+1)\mu_{2n}$,
so that \Eq{alphandef} can then be written as
\begin{eqnarray}
\alpha_n
&=&
\frac{\mu_{2n}-(2n-1)!!\mu_2^n}{(2n-1)!!\mu_2^n}.
\label{alphan}
\end{eqnarray}
We now see that even though both the $\alpha_n$ and the cumulants
$\kappa_{n>2}$ are, by construction, zero for Gaussian distributed
variables, in \Eq{alphan} the $\alpha_n$ are $2n$-th moments
$\mu_{2n}$ with only the most factored term, $\mu_2^n$, subtracted,
while the cumulants $\kappa_{2n}$ in \Eq{kappaintermsofmu} have all
possible factored terms subtracted.  Using \Eq{alphan} and the inverse
of the relation between $\kappa_n$ and $\mu_n$ in
\Eq{kappaintermsofmu} (which can be found by using the generating
functions), it is possible to express the $\alpha_n$ in terms of
$\kappa_n$ as
\begin{equation}
  \alpha_n \:=\: n!
\mathop{\sum_{0\leq p_\ell < n}}_{\sum_{\ell=1}^\infty  \ell p_\ell = n}
\prod_{\ell=1}^\infty\:\:
\left[
\frac{1}{p_\ell!}\left(\frac{2^\ell\kappa_{2\ell}}{(2\ell)!\kappa_2^\ell}\right)^{p_\ell}
\right]
.
\label{alphanintermsofkappan}
\end{equation}
According to this formal relation, the first few $\alpha_n$ are given by
\begin{subequations}
\label{43}
\begin{eqnarray}
\alpha_2 &=&\frac13 \frac{\kappa_4}{\kappa_2^2}
\label{Rahmansalpha2}
\\
\alpha_3 &=&\frac{\kappa_4}{\kappa_2^2}+\frac1{15}\frac{\kappa_6}{\kappa_2^3}
\end{eqnarray}
\end{subequations}

The Theorem in \Eq{Theorem0} says that for small times $t$,
$\kappa_2=\order{t^2}$ and $\kappa_{n>2}=\order{t^{2n}}$, so that
$\kappa_{2n}/\kappa_2^n=\order{t^{2n}}$. According to
\Eq{alphanintermsofkappan}, all $\alpha_n$ have a contribution from
$\kappa_{4}/\kappa_2^2$, so that all $\alpha_n$ are of $\order{t^4}$,
in contrast to the cumulants $\kappa_{2n}$, which are of increasing
order in $t$ with increasing $n$. In fact, using
\Eq{alphanintermsofkappan} and the Theorem, one can derive
straightforwardly that the dominant term for small $t$ in
\Eq{alphanintermsofkappan} is the one with $p_1=n-2$, $p_2=1$ and
$p_{\ell>2}=0$, which leads to
\begin{equation}
  \alpha_n \sim \frac{n(n-1)}{2}\,\alpha_2
\label{alphanrel}
\end{equation}
plus a correction of $\order{t^6}$.  Thus, for small $t$, $\alpha_3$ is
approximately three times $\alpha_2$, $\alpha_4$ six times
$\alpha_2$ etc.  Such approximate relations are indeed borne out by
Rahman's data on $\alpha_2$, $\alpha_3$ and
$\alpha_4$ for small $t$\cite{Rahman64,BoonYip80}.

In terms of the $\alpha_n$, the expansion of $G_s$ in
\Eq{afewgoodterms} becomes, with the help of \Eq{43},
\begin{equation}
  G_s(r,t) = \frac{e^{-w^2}}{\sqrt{2\pi\kappa_2}}\left[
1+\frac{3\alpha_2}{4!4}H_4(w)+\frac{15(\alpha_3-3\alpha_2)}{6!8}H_6(w)+\ldots
\right].
\label{afewgoodalphaterms}
\end{equation}
Although formally equivalent to \Eq{afewgoodterms}, for small $t$,
\Eq{afewgoodalphaterms} is somewhat less convenient, because one
cannot see right away that the last term (involving $H_6$) is
$\order{t^6}$ rather than $\order{t^4}$, as one may naively suspect
from $\alpha_2=\order{t^4}$ and $\alpha_3=\order{t^4}$.  Thus, a
cancellation between the leading orders of $\alpha_3$ and $3\alpha_2$
has to take place, which, especially using a numerical evaluation of
the $\alpha_n$, may be hard to obtain. On the other hand, in
\Eq{afewgoodterms}, this cancellation is automatic.

We note that, somewhat surprisingly, the data of Rahman also show that the
relation between $\alpha_n$ and $\alpha_2$ in \Eq{alphanrel} is still
approximately satisfied for larger times\cite{Rahman64}. This 
seems even true for hard spheres\cite{Desai66}. Thus the higher order
non-Gaussian parameters $\alpha_{n>2}$ are apparently dominated by
$\alpha_2$ for larger times just as they are for smaller times
$t$. This dominance of $\alpha_2$ makes it hard to extract from these
higher order non-Gaussian parameters any information that was not
already contained in $\alpha_2$. The cumulants $\kappa_n$, or perhaps
the $b_n$, may contain additional information about correlations in
(supercooled) fluids in a more accessible form (compared to the
$\alpha_n$), and may therefore be a more suitable choice to
investigate such correlations for all times $t$.

\begin{figure}[bt]
\graphfile{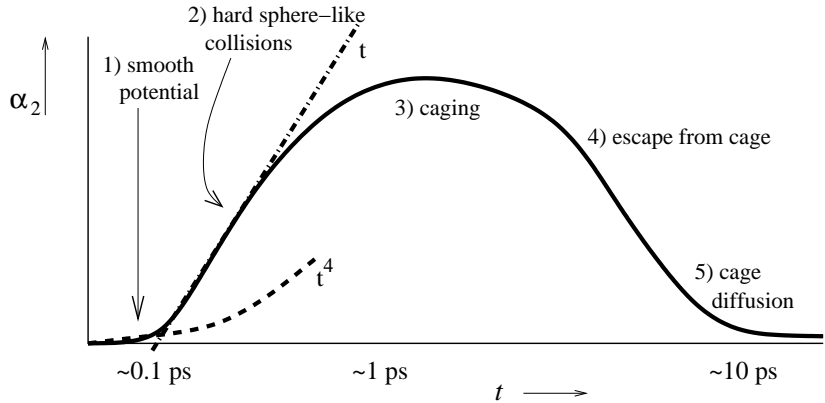}
\caption{Sketch of the behavior of the non-Gaussian parameter
$\alpha_2=\kappa_4/(3\kappa_2^2)$ for regular liquids in equilibrium
in different time regimes (based on fig.~7 in
ref.~\onlinecite{Rahman64} and fig.~4.11 in
ref.~\onlinecite{BoonYip80}), with our physical interpretation for
each regime.}
\end{figure}

c) Returning to the series in \Eq{equilGs}, although it may be
well-behaved for small enough $t$, it is not known up to what time
this remains so. Since $b_n=\order{t^{2n}}$ only to leading order in
$t$, there is a point in time after which the $\order{t^{2n}}$ term is
no longer a good approximation, and this may be related to the point
at which the series in \Eq{equilGs} is no longer guaranteed to be
useful. Rahman investigated $\alpha_2$ (as well as $\alpha_3$ and
$\alpha_4$) numerically for a model of liquid argon (temperature 94~K,
density $1.4\cdot 10^{3}$ kg/m$^{3}$)\cite{Rahman64}. (See also
ref.~\onlinecite{BoonYip80} for a broader overview.)  Figure~1 shows a
sketch of the non-Gaussian parameter $\alpha_2$ as a function of $t$,
based on fig.~7 in ref.~\onlinecite{Rahman64} and fig.~4.11 in
ref.~\onlinecite{BoonYip80}.  One sees that $\alpha_2$ is a very flat
function near $t=0$, which persists only up to roughly $t\approx
0.1$--$0.2$~ps. At that point the curve shoots up rapidly, leading to a
large ``hump'', which lasts up to about 10 ps (or perhaps somewhat below that),
after which it starts to decrease to zero.

Although somewhat outside the scope of this paper, we would like to
give a possible interpretation of the numerical results sketched
in Figure~1 for moderate and high densities. \emph{1)}~The flat
behavior of $\alpha_2$ near $t=0$ corresponds to the $\order{t^4}$
behavior as given by the Theorem in \Eq{Theorem0}. \emph{2)}~Because
hard spheres can be seen as a limit of a smooth interparticle
potential in which the steepness goes to infinity\cite{Sears72}, and
the potential used by Rahman is rather steep, the shoot-up phenomenon
at $\approx 0.1$~ps may well be related to the hard-spheres result of
De Schepper et al.\cite{DeSchepperetal81} that
$\kappa_{n}=\order{|t|^{n+1}}$, or
$\alpha_2=\order{|t|^5/t^4}=\order{|t|}$, as follows.  The steep but
smooth potential of Rahman will resemble a hard sphere fluid on time
scales $t_s$ on which a collision has been
completed\cite{DeSchepperCohen76}. Thus at $t=t_s$ the scaling
$\order{t^4}$ for $\alpha_2$ ought to go over to $\order{|t|}$, or in
fact, because of the duration $t_s$ of the `collision', to
$\order{|t-t_s|}$, which would require the kind of sharp increase
observed by Rahman at $\approx 0.1$ ps.  \emph{3)}~ While a
persistence of non-Gaussianity occurs already in dilute
systems\cite{Desai66}, this is due to a different mechanism than in
denser systems, where it comes about because the particle is trapped
in a ``cage'' formed by its neighboring particles, with which it has
repeated and correlated collisions. This effect is dominant at
high densities\cite{KobAndersen95}. \emph{4)}~The decay of $\alpha_2$
to zero indicates that the motion becomes Gaussian and presumably sets
in (for dense systems) when the particle manages to escape its
cage. After escaping, it finds itself in a new cage environment
consisting largely of particles with which it has not interacted
before.  This motion from cage to cage is called cage
diffusion\cite{CohenDeSchepper91}. \emph{5)}~From a central limit
theorem argument using that successively visited cages after many cage
escapes have little correlation with each other, one would then expect
Gaussian (and presumably but not necessarily diffusive) behavior.
The precise mechanism of the long-time behavior at low densities falls
outside the scope of this paper.

Also in simulations of a supercooled argon-like
mixture\cite{KobAndersen95}, $\alpha_2$ plotted as a function of time
shows a flat curve for short times and a sharp increase around $0.1$
ps, while $\alpha_3(t)$ shows similar behavior. The interesting part
from the perspective of supercooled liquids and glasses, however, is
in how far that increase continues and on what time scale and how
$\alpha_2$ decays back to zero, which takes a very long time for
supercooled liquids and is related to the time scale at which 
particles escape their cages. But the Theorem in
\Eq{phystheorem}--\Eq{Theorem0} has nothing to say about $\alpha_2$ on
that time scale.

\subsubsection{Local equilibrium systems}

As a first example of a non-equilibrium system to which the theorem
may apply, we consider a fluid not too far from equilibrium, such that
it has initially roughly an equilibrium distribution except that the
temperature, fluid velocity and density are spatially dependent, i.e.,
it is in local equilibrium. The class of initial distributions in
\Eq{physicaldistribution} does not seem to be of that form, and
indeed, if $\beta_{i}$ and $\mathbf u_i$ are allowed to vary with
$\mathbf r_i$, then the proof as given in the next part of this paper,
runs into difficulties. Nonetheless, one can construct distributions
of the form \eq{physicaldistribution} which physically describe
precisely the local equilibrium situation. Imagine dividing the
physical volume $\mathcal V$ into $M$ cells of equal size and
assigning to each cell $a$ a temperature $\beta_a$, a fluid velocity
$\mathbf u_a$ and a density $n_a$. The particles of the system are
divided up as well, putting $N_a=n_a \mathcal V/M$ particles in each
cell, such that particles $1$ through $N_1$ are in cell $1$, $N_1+1$
through $N_1+N_2$ are in cell $2$ etc. This can be accomplished by
choosing $f(\mathbf r^N)$ in \Eq{physicaldistribution} such that the
chance for these particles to be outside their cell is zero. Next, we
set all the $\beta_i$ and $\mathbf u_i$ of the particles in cell $a$
equal to $\beta_a$ and $\mathbf u_a$. If the cells are big enough so
that fluctuations in the number of particles can be neglected, this
situation describes local equilibrium just as well as spatially
dependent $\beta(\mathbf r_i)$ and $\mathbf u(\mathbf r_i)$ can, and
for this constructed local equilibrium, the Theorem in
\Eq{phystheorem} holds.

\subsubsection{Out-of-equilibrium phenomena on very short time scales}

In this section, we consider somewhat more general out-of-equilibrium
systems, namely those that start at $t=0$ with a Gaussian velocity
distribution. The distribution of the positions of the particles is
allowed to be very different from that in equilibrium, however.

The cumulants whose short time scaling was obtained here also occur
naturally in the Green's function theory, which was developed for
far-from-equilibrium phenomena on the picosecond and nanometer
scales\cite{Kincaid95,KincaidCohen02a,KincaidCohen02b,CohenKincaid02,VanZonCohen05b}.
Considering for example a mixture of two components, one can write the
density of component $\lambda$ ($\lambda=1$ or $2$) at position $x$
and for simplicity here in one dimension as:\cite{VanZonCohen05b}
\begin{equation}
    n_\lambda(x, t) = \int\! dx'\,G_\lambda(x,x',t)n_\lambda(x',0),
\end{equation}
where the non-equilibrium Green's function $G_\lambda(x,x',t)$
is the probability that a particle of component $\lambda$ is at
position $x$ at time $t$ given that it was at position $x'$ at time
zero. By an expansion detailed in a future
publication\cite{VanZonCohen05b} $G_\lambda$ can be written similarly as
$G_s$ in \Eq{equilGs}, as
\begin{equation}
\label{GFexpand}
     G_\lambda(x,x',t)=
\frac{e^{-w^2}}{\sqrt{2\pi \kappa_2}}.
\left[
       1 + \sum_{n=3}^{\infty} \frac{b_n}{(2\kappa_2)^{n/2}}
       H_n(w)
       \right] 
\end{equation}
Here the dimensionless $w\equiv(x-x'-\kappa_1)/\sqrt{2\kappa_2}$ and
\begin{equation}
  b_n \:=\: 
\mathop{\sum_{\{p_\ell\geq 0\}}}_{\sum_{\ell=3}^\infty \ell p_\ell = n}
\prod_{\ell=3}^\infty\:\:
\left[
\frac{1}{p_\ell!}\left(\frac{\kappa_\ell}{\ell!}\right)^{p_\ell}
\right]
.
\label{bnintermsofkappan}
\end{equation}
Some examples are: $b_3=\kappa_3/3!$, $b_4=\kappa_4/4!$,
$b_5=\kappa_5/5!$ and $b_6=(\kappa_6+10\kappa_3^2)/6!$.  These
equations are non-equilibrium generalizations of the equations for the
equilibrium Van Hove self-correlation function in \eq{equilGs} and
\Eqs{equilbnintermsofkappan}, respectively.  However, different from
the case of the Van Hove self-function, in the Green's function
theory, the $\kappa_n$, and thus the $b_n$ through
\Eq{bnintermsofkappan}, depend on $x'$ because in that theory the
single particle $i$ of component $\lambda$ is required to have been at
the position $x'$ at time zero. This requirement can be imposed by
multiplying the probability distribution function in
\Eq{physicaldistribution} by $\delta(x_i-x')$ (times a proper
normalization). The resulting distribution describes the subensemble
of the original ensemble for which particle $i$ is at $x'$ at time
zero. Note also that it is still of the same form as
\Eq{physicaldistribution}, with $f(\mathbf r^N)\rightarrow f'(\mathbf
r^N)=\delta(x_i-x')f(\mathbf r^N)$, where $x'$ is just a parameter.
Thus, as long as we restrict ourselves to systems with Gaussian
initial velocity distributions, the Theorem in \Eq{Theorem0} still
applies. Note that the $f(\mathbf r^N)$ may describe any
non-equilibrium distribution of the positions of the particles.

The Gaussian factor in \Eq{GFexpand} suggests as before that typical
values of $w$ are $\order{1}$, so $H_n(w)=\order{1}$. Its prefactor in
\Eq{GFexpand} is $b_n/(2\kappa_2)^{n/2}$. Using the Theorem in
\Eq{Theorem0} that $\kappa_n=\order{t^{2n}}$ and the relation
between $b_n$ and $\kappa_n$ in \Eq{bnintermsofkappan} it is easy to
see that also $b_{n}=\order{t^{2n}}$. Since $\kappa_2=\order{t^2}$, we
obtain
\begin{equation}
    \frac{b_n}{(2\kappa_2)^{n/2}}  = \order{t^n}.
\end{equation}
\emph{This means that, for these systems, the series for the
non-equilibrium Green's functions $G_\lambda$ in \Eq{GFexpand} are
well-behaved for small times $t$ (just as the equilibrium Van Hove
self-correlation function was) and that by truncating the series one
obtains for small $t$ approximations which can be systematically
improved by taking more terms into account.} This statement is in fact
not restricted to the one-dimensional case of the Green's functions
discussed here; the three-dimensional version, for which the $b_n$
become tensors and which will be presented
elsewhere\cite{VanZonCohen05b}, exhibits --- when the general
formulation in \Eq{phystheorem} is used --- the same scaling with $t$
of the subsequent terms in the series for the Green's functions.

We note that in non-equilibrium situations,
$\alpha_2=\kappa_4/(3\kappa_2^2)$ may have a similar behavior as
sketched in Figure 1 for the equilibrium $\alpha_2$. Although a proper
numerical test is yet to be performed, this expectation is roughly
consistent with numerical results of the Green's function for heat
transport\cite{KincaidCohen02a,KincaidCohen02b,CohenKincaid02}. In
that case the contribution of the non-Gaussian corrections in
\Eq{GFexpand} (involving the Hermite polynomials) were most
significant on the sub-picosecond time scale, whereas extrapolation
indicated that hydrodynamic-like results may occur for times possibly
as short as 2~ps\cite{KincaidCohen02b}.

In view of the possible application to nano-technology, it is
important to understand the behavior on very short time scales and the
related small length scales over which a particle typically moves at
such time scales. The Green's function theory can potentially describe
a system on all time scales. The current Theorem assures that this
theory can at least consistently describe the very short time
scales, by showing that the expansion of the Green's function is
well-behaved. For practical applications, and to know how short the
time scales must be for the Theorem to apply, it is still necessary to
determine the coefficient of the $\order{t^{2n}}$ of $\kappa_n$, i.e.\
the $c_n$ in \Eq{Theorem0} (cf.~\Subsec{coefficients}). This will
require a numerical evaluation of the moments of derivatives of the
forces, which we plan to do in the future. The behavior of $\kappa_n$
at longer time scales, will also be investigated in the future, with
special attention to the question whether the Green's functions give
hydrodynamic behavior for long times\cite{VanZonCohen05b}.

\section{Mathematical Part}

\label{mathematics}

\setcounter{equation}{0}

In this part, we present (\Subsec{thetheorem}) and prove
(\Subsec{proof}) the Theorem in its most general, mathematical form.

In the mathematical formulation of the Theorem, as well as in its
proof, it is convenient to adopt a different notation than the one
introduced in \Sec{physics}.  Since from a mathematical point of view
it does not matter whether different degrees of freedom are associated
with different spatial directions or with different particles, the
mathematical notation will treat all these degrees of freedom on the
same footing.  This can be achieved by calling the $x$ component of
particle one the first degree of freedom, its $y$ component the
second, its $z$ component the third, and then the $x$ component of
particle two the fourth degree of freedom, etc. The total number of
degrees of freedom is then $\mathcal N=3N$.

One can achieve the goal of describing all degrees of freedom in the
same way by associating with each degree of freedom a (generalized)
position $r_i$ and a (generalized) velocity $v_i$, and rewriting the
three dimensional real positions $\mathbf r_i$ and velocities $\mathbf
v_i$ by using the mapping
\begin{eqnarray}
\mathbf r_1\to\begin{pmatrix}r_1\\r_2\\r_3\end{pmatrix}, &
\mathbf r_2\to\begin{pmatrix}r_4\\r_5\\r_6\end{pmatrix},&
\ldots
\end{eqnarray}
and similar for the velocities and the average velocities, i.e.,
\begin{eqnarray}
\mathbf v_1\to\begin{pmatrix}v_1\\v_2\\v_3\end{pmatrix}, &
\mathbf v_2\to\begin{pmatrix}v_4\\v_5\\v_6\end{pmatrix},&
\ldots
\\
\mathbf v_1\to\begin{pmatrix}v_1\\v_2\\v_3\end{pmatrix}, &
\mathbf v_2\to\begin{pmatrix}v_4\\v_5\\v_6\end{pmatrix},&
\ldots
\end{eqnarray}
respectively.

Applying this mapping on the equations of motion \Eqs{eqmotion1}
and \eq{eqmotion2}, one finds that the generalized positions $r_i$
and velocities $v_i$ (collectively denoted by $r^\mathcal N$ and $v^\mathcal N$
respectively) satisfy 
\begin{eqnarray}
   \dot{r}_i &=& v_i
   \label{emotion1}
   \\
   \dot{v}_i &=& a_i(r^\mathcal N,t)
   \label{emotion2}
\end{eqnarray}
where the $a_i(r^\mathcal N,t)$ are the accelerations which follow from the mapping
\begin{eqnarray}
\frac{\mathbf F_1(\mathbf r^N,t)}{m_1}\to
\begin{pmatrix}a_1(r^\mathcal N,t)\\a_2(r^\mathcal N,t)\\a_3(r^\mathcal N,t)\end{pmatrix}, &
&\frac{\mathbf F_2(\mathbf r^N,t)}{m_2}\to
\begin{pmatrix}a_4(r^\mathcal N,t)\\a_5(r^\mathcal N,t)\\a_6(r^\mathcal N,t)\end{pmatrix}, \quad \ldots
\end{eqnarray}

We note that a possible interpretation of this formulation is that of
$\mathcal N$ ``particles'' moving in just one spatial dimension. This
convenient picture, in which each $r_i$ and $v_i$ may be seen as the
one-dimensional position and velocity of a ``particle'' $i$
respectively, will be used below, even though the system is not really
one-dimensional.  Cumulants that in the physical formulation of the
previous part involved both different spatial directions as well as
different particles (as e.g.\ on the lhs\ of \Eq{phystheorem}),
become in this picture cumulants involving just different ``particles''
(degrees of freedom), and these kinds of cumulants are therefore the
quantities of interest here.

We remark that the proof of the general Theorem that we will give
below is a ``physicists' proof'', meaning that the proof does not
claim to have complete mathematical rigor but has every appearance of
being correct, perhaps under mild and reasonable additional conditions
such as the existence of the moments and cumulants.

First, however, we need to introduce the general definition of
multivariate cumulants\cite{Cramer46,VanKampen}, needed in the Theorem
and its proof, and some of their properties.

\subsection{Preliminaries: the general definition and properties of cumulants}
\label{general definition}

Here we will give the general definition of cumulants which applies to
any number of variables, i.e., of the \emph{multivariate}
cumulants\cite{Cramer46,VanKampen}. Thereto we introduce a set of
(general) random variables $A_q$ with $q=1\ldots Q$, for which we
assume that a probability distribution function exists. The cumulant
generating function of the $A_q$ is defined as
\begin{equation}
  \Phi(k_1,\ldots k_Q) \equiv
 \log\baverage{\exp
  \sum_{q=1}^Q \mathrm{i} k_q A_q}.
\label{GXdef}
\end{equation}
Here $\average{\,}$ denotes an average with the probability
distribution function of the $A_q$. Note that if the $A_q$ are
expressed in terms of yet other random variables (later, in the
Theorem, the $r^\mathcal N$ and $v^\mathcal N$) the average may
equivalently be taken with their probability distribution function.
With the help of this generating function, the cumulants can be
expressed similarly as in \Eq{kappan} by
\begin{equation}
\cumulant{A_1^{[n_1]} ; \ldots ; A_Q^{[n_Q]}} 
\equiv
\left(
\prod_{q=1}^Q \dd[n_q]{\ }{(\mathrm{i} k_q)}\right)
\Phi(k_1,\ldots k_Q)\Bigg|_{\{k_q\}=0}.
\label{defcumvar}
\end{equation}
Note that $\{k_q\}=0$ is a short-hand notation for $k_1=0$, $k_2=0$,
\dots, $k_Q=0$. Furthermore, when one of the $n_q$ is equal to
$1$, we will omit the corresponding superscript $[1]$, e.g.,
$\cumulant{A_1;A_2}=\cumulant{A_1^{[1]};A_2^{[1]}}$. We stress once
more that our notation is different from Van Kampen's\cite{VanKampen},
who writes $\cumulant{A_1^{n_1}A_2^{n_2}\ldots A_Q^{n_Q}}$. In
particular, the square brackets in the superscripts are intended to
show that they are not powers of the $A_q$, while the semi-colons in
the expressions inside the double angular brackets indicate that one
should not interpret them as products, but as defined by
\Eq{defcumvar}. Note also that the expression $A_q^{[n_q]}$ only has a
meaning inside a cumulant.

The cumulants can be expressed in terms of moments 
analogously to \Eq{kappaintermsofmu}:
\begin{multline}
\cumulant{A_1^{[n_1]} ; \ldots ; A_Q^{[n_Q]}}
\\ =
- n_1!\cdots n_Q!\!
\mathop{\sum
_{\{p_{\{\ell\}}\geq 0\}}}
_{\sum_{\{\ell\}} \ell_q p_{\{\ell\}} = n_q}
\Big(\sum_{\{\ell\}} p_{\{\ell\}}-1\Big)!
\prod_{\{\ell\}} \frac{1}{p_{\{\ell\}}!}
\bigg(-\frac{\average{A_1^{\ell_1}\ldots A_Q^{\ell_Q}}}{\ell_1!\cdots\ell_Q!}\bigg)^{p_{\{\ell\}}}\!\!,
\label{cumintermsofmom}
\end{multline}
where $\{\ell\}=\{\ell_1,\ldots,\ell_Q\}$ denotes a set of $Q$
nonnegative $\ell_q$ values, $q=1\ldots Q$ and $p_{\{\ell\}}$ gives
the frequency of occurrence of that set. In \Eq{cumintermsofmom} the
sum is over the frequencies $p_{\{\ell\}}$ of all possible sets
$\{\ell\}$ (and thus all possible moments $\average{A_1^{\ell_1}\ldots
A_Q^{\ell_Q}}$) for which $\sum_{\ell=1}^\infty \ell_q p_{\{\ell\}} =
n_q$. This is nothing else than factorizing the expression
$A_1^{n_1}\ldots A_Q^{n_Q}$ in all possible ways. Hence, the cumulants
$\cumulant{A_1^{[n_1]};\ldots;
A_Q^{[n_Q]}}=\average{A_1^{n_1}\ldots A_Q^{n_Q}}\pm$ factored
terms that are products of the moments $\average{A_1^{\ell_1}\ldots
A_Q^{\ell_Q}}$ such that the $\ell_q$ values for fixed $q$ add up
(taking into account their frequencies $p_{\{\ell\}}$) to $n_q$.  For
some examples of multivariate cumulants, see
\Eqs{cumulantex1}-\eq{cumulantex3}.

We furthermore define the \emph{order} $n$ of a cumulant as the sum
$n=\sum_{q=1}^Qn_q$. Note that the maximum number of moments that are
multiplied in any term in \Eq{cumintermsofmom} is (taking into account
the frequencies) equal to the order of the cumulant.

We will now give four properties of the cumulants that follow from its
definition in  \Eqs{GXdef} and \eq{defcumvar} and that can be used to
manipulate expressions inside the double angular brackets of the cumulants.

\noindent
(a) The first property is that, as with averages, constants
may be taken in front of cumulants:
\begin{equation}
 \cumulant{(CA_1)^{[n_1]}; \ldots} = 
 C^{n_1} \cumulant{A_1^{[n_1]};\ldots}
\end{equation}

\noindent
(b) The second property concerns cumulants where the same quantity occurs more
than once:
\begin{eqnarray}
  \cumulant{A_1^{[n_1]};A_1^{[n_2]};\ldots}
&=&
  \cumulant{A_1^{[n_1+n_2]};\ldots}
\label{example}
\end{eqnarray}

\noindent
(c) The third is a multinomial expansion:
\begin{eqnarray}
\cumulant{(A_1+\ldots+A_Q)^{[n]};\ldots}
&=&
\mathop{\sum_{\{k_q\}\geq 0}}_{\sum_{q=1}^Q k_q = n}
\frac{n!}{\prod_{q=1}^Q k_q!}
\cumulant{A_1^{[k_1]};\ldots;A_Q^{[k_Q]};\ldots}.
\nonumber\\\label{starstar}
\end{eqnarray}

\noindent
(d) Finally, cumulants have the property  that they are shift
invariant if their orders are larger than
one\cite{Cramer46,VanKampen}. E.g., if $C_1$ is a constant,
\begin{equation}
  \cumulant{(A_1+C_1)^{[n]};\ldots}
= \cumulant{A_1^{[n]};\ldots}.
\end{equation}
Only if the order of the cumulant is one,  does a shift have any effect:
\begin{equation}
   \cumulant{A_1+C_1} = \cumulant{A_1}+C_1.
\end{equation}

\subsection{The Theorem}
\label{thetheorem}

\begin{maintheorem}
For a classical dynamical system of $\mathcal{N}$ degrees of
freedom described by (generalized) coordinates $r_i$ and  velocities
$v_i$ ($i=1\ldots \mathcal N$), collectively denoted by $r^\mathcal N$
and $v^\mathcal N$ respectively, for which

\begin{enumerate}

\item[a)] the time evolution is given by the equations of motion
  \begin{eqnarray}
   \dot{r}_i &=& v_i
   \label{motion1}
   \\
   \dot{v}_i &=& a_i(r^\mathcal N,t)
   \label{motion2}
  \end{eqnarray}
  with velocity-independent, smooth accelerations $a_i$, and

\item[b)] the \emph{initial} ensemble is described by a probability
  distribution in which the velocities are 
  multivariate Gaussian\cite{VanKampen} and independent of the coordinates
  $r_i$, i.e., with a distribution function of the form
  \begin{equation}
   \mathcal P(r^\mathcal N,v^\mathcal N)=
   f(r^\mathcal N)
   \sqrt{\mathrm{det}\,[\mathsf \Xi/(2\pi)]}
    \exp\bigg[-\frac12\sum_{i,j=1}^\mathcal N
   \Xi_{ij}(v_i-u_i)(v_j-u_j)\bigg],  
   \label{distribution}
  \end{equation}
  where $\mathsf \Xi$ is a positive, symmetric $\mathcal N\times\mathcal N$
  matrix with constant elements $\Xi_{ij}$, $u_i$ is the average
 velocity corresponding to the $i$th degree of freedom, and $f(r^\mathcal N)$ is the
  probability distribution of the coordinates.
\end{enumerate}
When these conditions are satisfied, the cumulants\footnote{These
cumulants are defined in \Subsec{general definition} with $Q\to\mathcal N$,
$q\to i$ and $A_q\to\Delta r_i$} 
\begin{equation}
\kappa_{\{n_i\}}\equiv
\langle\!\langle \Delta r_1^{[n_1]}; \Delta r_2^{[n_2]}; \ldots;\Delta
r_\mathcal N^{[n_\mathcal N]} \rangle\!\rangle 
\end{equation}
of the displacements
\begin{equation}
  \Delta r_i \equiv
r_i(t)-r_i(0) 
\end{equation}
(whose $t$ dependence has been suppressed)
satisfy for sufficiently short initial times~$t$
\begin{equation}
\kappa_{\{n_i\}}
= \begin{cases}
               c_{\{n_i\}} t^{n}+ \order{t^{n+1}} &  \text{if }\: n\leq2
		\\
               c_{\{n_i\}} t^{2n}+ {\order{t^{2n+1}}} &   \text{if }\: n>2,
\end{cases}
\label{Generalized Theorem}
\end{equation}
where $n$ is the order of the cumulant given by
\begin{equation}
n=n_1+n_2+\ldots+n_\mathcal N.
\label{ndef}
\end{equation}
The coefficients
$c_{\{n_i\}}$ will be given later in \Eq{cn} in \Subsec{coefficients}.
\end{maintheorem}

\subsection{Proof of the Theorem}

\label{proof}

\subsubsection*{Strategy based on Gaussian velocities}

The Theorem formulated in Eqs.~(\ref{motion1}--\ref{ndef}) in
\Subsec{thetheorem} will be proved in this section, although we will
defer the details of the proof of a required auxiliary theorem to the
Appendix for greater clarity of the procedure.

To obtain the initial, short time behavior of the moments and
cumulants of the displacement, $\Delta r_i$ may be Taylor-MacLaurin
expanded around $t=0$ as
\begin{equation}
  \Delta r_i = 
\sum_{\gamma =1}^{\infty} 
  \frac{d^\gamma r_i}{dt^\gamma}\Big|_{t=0}
\,\frac{t^\gamma}{\gamma!} \:
=\sum_{\gamma=1}^\infty \ddh[\gamma-1]{v_i}{t}\Big|_{t=0}
\, \frac{t^\gamma}{\gamma!}
\label{expansion}
\end{equation}
where we used \eq{motion1}.
Because of the equations of motion
\eq{motion1} and \eq{motion2}, the $d^\gamma v_i/dt^\gamma$, viewed as
functions of $r^\mathcal N$ and $v^\mathcal N$ (at time $t$) as well as
explicitly of $t$, are recursively
related by
\begin{equation}
\ddh[\gamma+1]{v_i}{t}
= \sum_{j=1}^\mathcal N
\left[
v_j
\dd{\ }{r_j}\Big(\ddh[\gamma]{v_i}{t}\Big)
+
a_j(r^\mathcal N,t)
\dd{\ }{v_j}\Big(\ddh[\gamma]{v_i}{t}\Big)
\right]
+\dd{\ }{t}\Big(\ddh[\gamma]{v_i}{t}\Big)
\label{recursion}.
\end{equation}

To show that the cumulant $\kappa_{\{n_i\}}=\order{t^{2n}}$ for $n>2$,
it
is of course possible to work out this cumulant  straightforwardly  using
\Eqs{cumintermsofmom}, \eq{expansion}, \eq{recursion} and
\eq{distribution}, in that order.  Such a procedure was essentially
followed by Schofield\cite{Schofield61} for
$\kappa_n=\kappa_{\{n,0,0,\ldots\}}=\cumulant{\Delta r^{[n]}_1}$ for
$n\leq6$ and Sears\cite{Sears72} for $n\leq8$ for equilibrium
fluids. In their expressions many cancellations occurred before
$\kappa_n$ could be seen to be, for these cases, of \textorder{t}{2n}
instead of \textorder{t}{n}.  These cancellations seemed to happen as
a consequence of equilibrium properties.  However, by carrying out the
same straightforward procedure for the more general class of
non-equilibrium initial conditions in \Eq{distribution}, we have found
that, while odd moments are no longer zero, still
$\kappa_n=\order{t^{2n}}$ for $n=3$, $4$, $5$ and $6$.  These results
for $\kappa_3$, $\kappa_4$, $\kappa_5$ and $\kappa_6$ naturally led us
to propose the Theorem.  Because the straightforward calculations for
this non-equilibrium case are very lengthy, they will not be presented
here. In any case this procedure is not very suited to determine the
order in $t$ of $\kappa_{\{n_i\}}$ for general $n_i$, because with
increasing order $n=\sum_{i=1}^\mathcal N n_i$ an increasing number of
terms have to be combined (taking together equal powers of $t$ from
the various products of moments) before they can be shown to be zero.

\emph{Our strategy for proving that  $n$th order cumulants
$\kappa_{\{n_i\}}=\order{t^{2n}}$ for all $n>2$, will be to
exploit the Gaussian distribution of the velocities as much as
possible.}  But we can only hope to use the Gaussian nature of the
velocities if we succeed in bringing out explicitly the dependence of
the coefficients of the power series in $t$ of the cumulants on the
velocities. This dependence has so far only been given implicitly ---
the cumulants are related to the moments by \Eq{cumintermsofmom}, the
moments contain $[\Delta r_i]^\ell$, $\Delta r_i$ is expanded in the
time $t$ in \Eq{expansion}, and the coefficients in that expansion are
the derivatives of the velocity $v_i$, which can be found by using
\Eq{recursion} recursively.  To make this more explicit, the first
step\ of the proof will be to expand the cumulants as a powers series
in the time $t$ and the second step\ will be to express this series
more explicitly in the velocities.  In the third and last step\
we will then use the properties of Gaussian distributed variables,
i.e., the velocities, to complete the proof of the Theorem.

It turns out that the properties of Gaussian distributed variables
that we will require in the third step\ of the proof are
formulated for \emph{independent} Gaussian variables whose \emph{mean
is zero}, while in \Eq{distribution} the velocities are Gaussian but
do not have zero mean, nor are they independent (because $\Xi_{ij}\neq0$
for $i\neq j$
in general).  For this reason, it is convenient to introduce already
at this point new velocity variables whose mean is zero
(cf. \Eq{distribution}):
\begin{equation}
  V_i \equiv v_i-\average{v_i} = v_i-u_i.
\label{Vdef}
\end{equation}
The $V_i$ are generalizations of the peculiar velocities used in
kinetic theory and will be referred to in the general case treated in
this paper as \emph{peculiar velocities} as well.  For the same
reason, it is convenient to get rid of the statistical dependence of
the initial velocities in \Eq{distribution} by bringing the (positive)
matrix $\mathsf \Xi$ to its diagonal form $\xi_i\delta_{ij}$. This can
be accomplished by an orthogonal transformation. We will assume here
that this orthogonal transformation has been performed so that $\mathsf \Xi$ is
diagonal, and will show at the end of the proof that the
form~\Eq{Generalized Theorem} of the Theorem is invariant under such a
transformation (while the coefficients $c_{\{n_i\}}$ do change).

Substituting \Eq{Vdef} into the time expansion in \eq{expansion} and
into the probability distribution \Eq{distribution} (using that
$\mathsf \Xi$ is now diagonal), gives the expansion of $\Delta
r_i$ and the probability distribution $\mathcal P(r^\mathcal
N,V^\mathcal N)$ in their peculiar velocity
form:
\begin{eqnarray}
\Delta r_i
 &
=& 
u_i t +
\sum_{\gamma=1}^{\infty} \frac{t^\gamma}{\gamma!} \ddh[\gamma-1]{V_i}{t}\Big|_{t=0}.
\label{neexpansion}
\end{eqnarray}
\begin{eqnarray}
\mathcal P(r^\mathcal N,V^\mathcal N)&=&
  f(r^\mathcal N)\prod_{i=1}^\mathcal N
\sqrt{2\pi \xi_i}\,\exp[-\mbox{$\frac12$}\xi_i V_i^2]
\label{newdistribution}
\end{eqnarray}
respectively.
In order to make future expressions less complicated we introduce for
the coefficients in \Eq{neexpansion} the notation ($\gamma=1,$ $2,$
$\ldots$)
\begin{equation}
  X_{i\gamma}(r^\mathcal N,V^\mathcal N) 
=\frac{1}{\gamma!} \ddh[\gamma-1]{V_i}{t}\Big|_{t=0}.
\label{wdef}
\end{equation}
Thus \Eq{neexpansion} becomes
\begin{equation}
\Delta r_i
=
u_i t +
\sum_{\gamma=1}^{\infty} X_{i\gamma}(r^\mathcal N,V^\mathcal N) \,t^\gamma,
\label{newexpansion}
\end{equation}
We remark here that in the double indices of
$X_{i\gamma}(r^\mathcal N,V^\mathcal N)$, the first index always pertains to a
particle number and will be denoted by the roman letter $i$, while the
second pertains to an order in $t$ and will be denoted by the Greek
letter $\gamma$.  Below, we will drop the explicit dependence of the
$X_{i\gamma}$ on $r^\mathcal N$ and $V^\mathcal N$.

We will now start the actual proof of the Theorem for general $n$.

\subsubsection*{First step: Expanding the cumulants in the time $t$}

The infinite number of terms in the time expansion $\Delta r_i$ in
\Eq{newexpansion} means that to get the power expansion in $t$ of the
cumulants $\kappa_{\{n_i\}}= \cumulant{\Delta
r_1^{[n_1]};\ldots;\Delta r_\mathcal N^{[n_{\mathcal N}]}} $ occurring
on the lhs\ of \Eq{Generalized Theorem}, we would use
\Eq{cumintermsofmom} with
\begin{equation}
Q\to\mathcal N,\quad
q\to i, \quad A_q\to\Delta r_i. 
\label{subst}
\end{equation}
and combine the infinite number of terms coming from \Eq{newexpansion}.
Obviously if we are interested in the cumulants up to
$\order{t^{2n-1}}$, where $n=\sum_{i=1}^\mathcal Nn_i$
(cf.~\Eq{ndef}), we should not have to retain all these terms in
\Eq{newexpansion}, but only those up to $\order{t^{2n-1}}$. As a
matter of fact, we need even less terms, namely only terms up to
$\order{t^n}$ in \Eq{newexpansion}, as the following reasoning shows.
The cumulants occurring on the lhs\ of \Eq{Generalized Theorem} are
given in terms of the moments $\average{\Delta r_1^{\ell_1}\Delta
r_2^{\ell_2}\ldots\Delta r_\mathcal N^{\ell_\mathcal
N}}=\average{\prod_{i=1}^\mathcal N\Delta r_i^{\ell_i}}$ by
\Eq{cumintermsofmom}. Taking the terms up to \textorder{t}{n} in
\Eq{newexpansion}, i.e., writing $\Delta r_i=u_i t +\sum_{\gamma=1}^n
X_{i\gamma}t^\gamma + \order{t^{n+1}}$, it is straightforward to show
that the moments satisfy
\begin{equation}
\baverage{\prod_{i=1}^\mathcal N\Delta r_i^{\ell_i}} \:=\:
\baverage{\prod_{i=1}^\mathcal N \big(u_i t+\sum_{\gamma=1}^n X_{i\gamma}
t^\gamma\big)^{\ell_i}}
+\order{t^{n+\sum_{i=1}^\mathcal N \ell_i}}
\end{equation}
so that, for given $p_{\{\ell\}}$,
\begin{equation}
\prod_{\{l\}}
\baverage{\prod_{i=1}^\mathcal N\Delta r_i^{\ell_i}}^{p_{\{\ell\}}} 
=
\prod_{\{l\}}
\baverage{\prod_{i=1}^\mathcal N \big(u_it+\sum_{\gamma=1}^n X_{i\gamma}
t^\gamma\big)^{\ell_i}}^{p_{\{\ell\}}}
+\order{t^{n+\sum_{\{\ell\}}p_{\{\ell\}}\sum_{i=1}^\mathcal N
    \ell_i}}.
\label{aotemp}
\end{equation}
Products of this kind occur in the definition of the cumulants on the
rhs\ of \Eq{cumintermsofmom}, and are summed over $p_{\{\ell\}}$ with the
restriction that (with $q=i$ here)
$\sum_{\{\ell\}}\ell_ip_{\{\ell\}}=n_i$. Since also
$\sum_{i=1}^\mathcal N n_i=n$, \Eq{aotemp} becomes
\begin{equation}
\prod_{\{l\}}
\baverage{\prod_{i=1}^\mathcal N\Delta r_i^{\ell_i}}^{p_{\{\ell\}}} \:=\:
\prod_{\{l\}}
\baverage{\prod_{i=1}^\mathcal N \big(u_it+\sum_{\gamma=1}^n X_{i\gamma}
t^\gamma\big)^{\ell_i}}^{p_{\{\ell\}}}
+\order{t^{2n}}.
\end{equation}
As this holds for each such expression on the rhs\ of
\Eq{cumintermsofmom} (with~\Eq{subst}), we have for its
lhs
\begin{eqnarray}
\kappa_{\{n_i\}} &=&
\cumulant{\Delta r_1^{[n_1]};\ldots;\Delta r_\mathcal N^{[n_\mathcal
      N]}
}
\nonumber\\&=&
\bcumulant{ \big(u_1t+\sum_{\gamma=1}^n X_{1\gamma}
t^\gamma\big)^{[n_1]}
;
\ldots
;
\big(u_\mathcal Nt+\sum_{\gamma=1}^n X_{\mathcal N\gamma}
t^\gamma\big)^{[n_\mathcal N]}}
\nonumber\\
&&+\order{t^{2n}}.
\label{lemmaA3}
\end{eqnarray}

We remark that the first term on the rhs\ of this equation gives all
powers $t^n$ up to $t^{2n-1}$, which we are interested in, as well as
some of the powers of $t$ higher than $2n$, which we are not interested
in. The second term, i.e., $\order{t^{2n}}$ only contains
powers of $t$ of $2n$ and higher.  So with \Eq{lemmaA3} we have
established that for the lower powers of $t$ only $n$ terms in the
time expansion of $\Delta r_i$ are needed, but we have not
separated the powers of $t$ lower and higher than $2n$ completely yet.

For that purpose we continue from \Eq{lemmaA3} and use first the
shift invariance property of cumulants explained at the end of
\Subsec{general definition}, to obtain
from \Eq{lemmaA3} 
\begin{eqnarray}
\kappa_{\{n_i\}}
&=&
\bcumulant{ \big(\sum_{\gamma=1}^n X_{1\gamma}
t^\gamma\big)^{[n_1]}
;
\ldots
;
\big(\sum_{\gamma=1}^n X_{\mathcal N\gamma}
t^\gamma\big)^{[n_\mathcal N]}}
\nonumber\\&&
+\sum_{i=1}^\mathcal N u_it\,\delta_{n_i1}\prod_{j\neq i}\delta_{n_j0}
+\order{t^{2n}}.
\label{an29}
\end{eqnarray}
Furthermore, using the multinomial expansion in \Eq{starstar}, we can
write
\begin{align}
&
\bcumulant{ \big(\sum_{\gamma=1}^n X_{1\gamma}
t^\gamma\big)^{[n_1]}
;
\ldots
;
\big(\sum_{\gamma=1}^n X_{\mathcal N\gamma}
t^\gamma\big)^{[n_\mathcal N]}}
\nonumber\\&
=
\mathop{\sum_{\{n_{i\gamma}\geq 0\}}}_{\sum_{\gamma=1}^n
    n_{i\gamma}=n_i}
\left[
\prod_{i=1}^\mathcal N\frac{n_i!}{\prod_{\gamma=1}^n n_{i\gamma}!}
\right]
\cumulant{X_{11}^{[n_{11}]};X_{12}^{[n_{12}]};\ldots;X_{\mathcal
 Nn}^{[n_{\mathcal N n}]}}
 t^{\sum_{i=1}^\mathcal N\sum_{\gamma=1}^n n_{i\gamma}\gamma}
\label{a32}
\end{align}
Note that the summation indices $n_{i\gamma}$ arise from the
multinomial expansion, where $i$ denotes a particle index and $\gamma$
runs from $1$ to $n$.  For $\kappa_{\{n_i\}}$ in \Eq{an29}, we need
this quantity only explicitly up to $\order{t^{2n-1}}$, so powers of
$t$ higher than $2n-1$ may be discarded (i.e., combined with the
$\order{t^{2n}}$ term) and only powers lower than $2n$ need to be
kept. Hence, in the exponent on the rhs\ of \Eq{a32}, we only need
terms with $\sum_{i=1}^\mathcal
N\sum_{\gamma=1}^nn_{i\gamma}\gamma<2n$.  Since also
$n=\sum_{i=1}^\mathcal N\sum_{\gamma=1}^nn_{i\gamma}$, we obtain
$\sum_{i=1}^\mathcal N\sum_{\gamma=1}^nn_{i\gamma}\gamma < 2
\sum_{i=1}^\mathcal N\sum_{\gamma=1}^nn_{i\gamma}$, or
$\sum_{i=1}^\mathcal N n_{i1}>\sum_{i=1}^\mathcal N\sum_{\gamma=2}^n
n_{i\gamma}(\gamma-2)$. Combining this condition with \Eqs{an29} and
\eq{a32}, and using $X_{i1}=V_i$ (cf.\ \Eq{wdef}), we find that the
expansion of an $n$th order cumulant in time $t$ up to
$\order{t^{2n-1}}$ is given by
\begin{multline}
\kappa_{\{n_i\}}=
\mathop{
\mathop{\sum_{\{n_{i\gamma}\geq 0\}}}_{\sum_{\gamma=1}^n
    n_{i\gamma}=n_i}
}
_{\sum_{i=1}^\mathcal N n_{i1}\:>\:\sum_{i=1}^\mathcal N \sum_{\gamma=2}^n  n_{i\gamma}(\gamma-2)}
\!\!\!\!\!\!\!
\left[
\prod_{i=1}^\mathcal N\frac{n_i!}{\prod_{\gamma=1}^n n_{i\gamma}!}
\right]
\\
\times
\cumulant{V_{1}^{[n_{11}]};\ldots;V_\mathcal
    N^{[n_{\mathcal N1}]}
; X_{12}^{[n_{12}]};
\ldots; X_{\mathcal N n}^{[n_{\mathcal Nn}]}}
 t^{\sum_{i=1}^\mathcal N\sum_{\gamma=1}^n n_{i\gamma}\gamma}
\\
+\sum_{i=1}^\mathcal N u_it\,\delta_{n_i1}\prod_{j\neq i}\delta_{n_{j}0}
+\order{t^{2n}}.
\label{expanded}
\end{multline}

We note that in \Eq{expanded} only cumulants appear, instead of
moments, and, more importantly, that powers of $t$ lower and higher
than $2n$ are easily identified, something that in the straightforward
moment-based approach mentioned above only happens after a lengthy
calculation.

\subsubsection*{Second step: Writing cumulants in terms of
  peculiar velocities}

The dependence of $\kappa_{\{n_i\}}$
in \Eq{expanded} on the peculiar velocities $V^\mathcal N$ follows
from the dependence of the $X_{i\gamma}$ on the $V^\mathcal N$.

According to \Eq{wdef}, the coefficients
$X_{i\gamma}$ can be determined using
\begin{equation}
X_{i\gamma}(r^\mathcal N,V^\mathcal N)= \tilde
X_{i\gamma}(r^\mathcal N,V^\mathcal N,t)\big|_{t=0},
\label{226p}
\end{equation}
where we have defined 
\begin{equation}
\tilde X_{i\gamma}(r^\mathcal N,V^\mathcal N,t)=\frac1{\gamma!}
\ddh[\gamma-1]{V_i}{t}
\label{226pp}
\end{equation}
taken at time $t$.  Below, we will suppress the explicit dependence of 
$\tilde X_{i\gamma}$  on $r^\mathcal N$, $V^\mathcal N$ and $t$.  The $\tilde
X_{i\gamma}$ satisfy the recursion relation, from \Eqs{recursion}:
\begin{eqnarray}
 \frac{\tilde X_{i\gamma+1}}{\gamma+1} &
=& 
 \sum_{j=1}^\mathcal N
\left[
(u_j+V_j)
\dd{\tilde X_{i\gamma}}{r_j}
+
a_j(r^\mathcal N,t)
\dd{\tilde X_{i\gamma}}{V_j}
\right]
+\dd{\tilde X_{i\gamma}}{t}
.
\label{newerrecursion}
\end{eqnarray}

For $\gamma=2$, \Eqs{226pp}, \eq{Vdef} and \eq{motion2} show that
$\tilde X_{i2}=(dV_i/dt)/2=a_i(r^\mathcal N,t)$, which is independent of
$V^\mathcal N$ and is thus a polynomial in the peculiar
velocities $V^\mathcal N$ of total degree zero. Using the recursion
relation \Eq{newerrecursion} one sees that if $\tilde X_{i\gamma}$ is a
polynomial in $V^\mathcal N$ of total degree $\gamma-2$ then on the
rhs\ of \Eq{newerrecursion}, the term
$(u_j+V_j)\partial{\tilde X_{i\gamma}}/\partial{r_j}$ is a polynomial of
total degree $\gamma-1$, while
$a_j\partial{\tilde X_{i\gamma}}/\partial{V_j}$ has a total degree of
$\gamma-3$, and $\partial{\tilde X_{i\gamma}}/\partial{t}$ has a total degree
$\gamma-2$. The highest total power of $V^\mathcal N$ in
$\tilde X_{i\gamma+1}$ is thus $\gamma-1$.

In other words, for $\gamma\geq2$, $\tilde X_{i\gamma}$ is a polynomial in
the peculiar velocities $V^\mathcal N$ of total degree $\gamma-2$,
with coefficients that can depend on the positions of the particles.
From \Eq{226p} we then see that also $X_{i\gamma}=\tilde X_{i\gamma}(t=0)$ is a polynomial in
the peculiar velocities $V^\mathcal N$ of total degree $\gamma-2$.
Therefore we can write for the dependence of $X_{i\gamma}$ on $V^\mathcal N$
\begin{eqnarray}
X_{i\gamma} &=& \sum_{p=0}^{\gamma-2} \sum_{\sum_j p_j=p}
b_{i\{p_j\}}(r^\mathcal N)\, V_1^{p_1}\cdots V_\mathcal N^{p_\mathcal
  N}
\label{Xpoly}
\end{eqnarray}
It turns out that in the third step\ of the proof we will not
need the precise and explicit forms of
the $b_{i\{p_j\}}(r^\mathcal N)$, but only that the \emph{total degree} of
the polynomial $X_{i\gamma}$ is $\gamma-2$.

\subsubsection*{Third step: Using the Gaussian nature of velocities}

Given the polynomial nature of the $X_{i\gamma}$ as a function of
$V^\mathcal N$ in \Eq{Xpoly}, we can give the following interpretation to
the second condition under the summation sign in \Eq{expanded}, i.e.,
$\sum_{i=1}^\mathcal N n_{i1}>\sum_{i=1}^\mathcal N\sum_{\gamma=2}^n
n_{i\gamma}(\gamma-2)$. 
Each expression of the form $X_{i\gamma}^{[n_{i\gamma}]}$ in the
cumulant $\cumulant{V_{1}^{[n_{11}]};\ldots;V_\mathcal N^{[n_{\mathcal
N1}]} ; X_{12}^{[n_{12}]}; \ldots; X_{\mathcal N n}^{[n_{\mathcal
Nn}]}}$ in \Eq{expanded} (although devoid of meaning outside of
cumulant brackets) signifies that $n_{i\gamma}$ is the highest
power of $X_{i\gamma}$ occuring inside the averages in the expression
for the cumulant in terms of moments on the right hand side of
\Eq{cumintermsofmom}. Because $X_{i\gamma}$ is a polynomial in
$V^\mathcal N$ of total degree $(\gamma-2)$, this highest power of
$X_{i\gamma}$ is a polynomial in $V^\mathcal N$ of total degree
$n_{i\gamma}(\gamma-2)$.
Then, all such expressions $X_{i\gamma}^{[n_{i\gamma}]}$ together in
the cumulant $\cumulant{V_{1}^{[n_{11}]};\ldots;V_\mathcal
N^{[n_{\mathcal N1}]} ; X_{12}^{[n_{12}]}; \ldots; X_{\mathcal N
n}^{[n_{\mathcal Nn}]}}$ in \Eq{expanded} generate polynomials of at most
degree $\sum_{\gamma=2}^nn_{i\gamma}(\gamma-2)$ inside the 
averages in the expression
for the cumulant in terms of moments on the right hand side of
\Eq{cumintermsofmom}.
On the other hand the expressions $V_i^{[n_{i1}]}$  in the cumulant
$\cumulant{V_{1}^{[n_{11}]};\ldots;V_\mathcal N^{[n_{\mathcal N1}]} ;
X_{12}^{[n_{12}]}; \ldots; X_{\mathcal N n}^{[n_{\mathcal Nn}]}}$ in
\Eq{expanded} together generate polynomials of at most degree
$\sum_{i=1}^\mathcal N n_{i1}$.
Thus, the condition  
$\sum_{i=1}^\mathcal N n_{i1}>\sum_{i=1}^\mathcal N\sum_{\gamma=2}^n
n_{i\gamma}(\gamma-2)$ in \Eq{expanded} indicates that the
total degree in $V^\mathcal N$ generated by the $V_i$ in the cumulant
$\cumulant{V_{1}^{[n_{11}]};\ldots;V_\mathcal N^{[n_{\mathcal N1}]} ;
X_{12}^{[n_{12}]}; \ldots; X_{\mathcal N n}^{[n_{\mathcal Nn}]}}$ 
is larger than the total degree generated by the
$X_{i\gamma}$.

The crucial point is now that the auxiliary \Theorem{cumulant with v
power} in the Appendix \label{AppAneeded} can be applied to cumulants
of this form.  This theorem is most conveniently expressed in terms of
general random variables $P_q$, which were used there as well to
denote general random variables, but which are now polynomial
functions of the peculiar velocities $V^\mathcal N$. \Theorem{cumulant
with v power} states that if $P_q$ ($q=1\ldots Q$) are polynomials of
total degree $d_q$ in the independent, zero-mean Gaussian distributed
$V^\mathcal N$, and $n_q$ and $p_i$ are nonnegative integers, then
$\cumulant{V_1^{[p_1]};\ldots; V_\mathcal N^{[p_\mathcal N]};
P_1^{[n_1]};\ldots; P_Q^{[n_Q]}}=0$ if $\sum_{i=1}^\mathcal N
p_{i}>\sum_{q=1}^Qn_q d_q$, except when all $n_{q\geq 1}=0$ and one
$p_i=2$, in which case it becomes just $\cumulant{V^{[2]}_i}$.

In order to apply \Theorem{cumulant with v power} to each term in the
summation on the rhs\ of \Eq{expanded}, we need to rewrite
$\cumulant{V_1^{[p_1]};\ldots; V_\mathcal N^{[p_\mathcal
N]};X_{12}^{[n_{12}]};\ldots;X_{\mathcal Nn}^{[n_{\mathcal Nn}]}}$ as
$\cumulant{V_1^{[p_1]};\ldots; V_\mathcal N^{[p_\mathcal
N]};P_1^{[n_1]};\ldots;P_Q^{[n_Q]}}$. This can be achieved by a
mapping of single to double indices, i.e., by setting
\begin{subequations}
\begin{equation}
Q\to\mathcal N(n-1)
\end{equation}
\begin{equation}
\begin{array}{r@{\,}c@{\,}l@{\!\!\!\!\!}r@{\,}c@{\,}l}
(P_1,n_1,d_1)&\to&(X_{12},n_{12},0)&
(P_n,n_n,d_n)&\to&(X_{22}, n_{22}, 0)
\\
(P_2,n_2,d_2)&\to&(X_{13}, n_{13}, 1)&
(P_{n+1},n_{n+1},d_{n+1})&\to&(X_{23}, n_{23}, 1)
\\
(P_3,n_3,d_3)&\to&(X_{14}, n_{14},2)&
&\vdots&
\\
&\vdots&&
&\vdots&
\\
(P_{n-1},n_{n-1},d_{n-1})&\to&(X_{1n}, n_{1n}, n-2)&
(P_Q,n_Q,d_Q)&\to&(X_{\mathcal N n}, n_{\mathcal N n}, n-2)
\end{array}
\end{equation}
and 
\begin{equation}
p_i\:\:\to\:\:n_{i1} 
\end{equation}
\end{subequations}
Then $\sum_{q=1}^Q n_qd_q$ is seen to be equal to
$\sum_{i=1}^\mathcal N\sum_{\gamma=1}^n n_{i\gamma}(\gamma-2)=d$. As the
restriction on the summation in \Eq{expanded} shows, for all terms on
the rhs\ of \Eq{expanded} (except the $\order{t^{2n}}$ of course),
$\sum_{i=1}^\mathcal N n_{i1}$ is larger than this
$d$. \Theorem{cumulant with v power} tells us that the cumulant
occurring in each term of these terms is then zero except when
$n_{i2}=n_{i3}=\cdots n_{in}=0$ for all $i$ and only one $n_{i1}=2$
and $n_{j\neq i1}=0$. This exception means, since also
$\sum_{i=1}^\mathcal N\sum_{\gamma=1}^n n_{i\gamma}=n$, that
$n=n_{i1}=2$. So the only possible nonzero term in {the sum in}
\Eq{expanded} occurs for $n=2$.  Furthermore, for $n=1$, the last term
$\sum_{i=1}^\mathcal N u_it\,\delta_{n_{i}1}\prod_{j\neq
i}\delta_{n_{j}0}$ in \Eq{expanded} is also left.

Thus, using \Theorem{cumulant with v power}, we have shown that each
term in \Eq{expanded} is zero separately except for $n=1$ and $n=2$,
so that
\begin{equation}
\kappa_{\{n_i\}}
 = \begin{cases}
                \order{t^n} &  \quad \text{if }\: n\leq2
\\
                \order{t^{2n}} & \quad \text{if }\: n> 2
             \end{cases}
\label{thisone}
\end{equation}
remains on the rhs\ of \Eq{expanded}. Given that $\kappa_{\{n_i\}}$ can
be expanded as a power series in $t$, \Eq{thisone} coincides with the
formulation \Eq{Generalized Theorem} of the Theorem, if we define
$c_{\{n_i\}}$ as the coefficients of the $t^{n}$ and $t^{2n}$,
respectively. This is therefore now proved (with the proviso that
Theorem A is proved in the Appendix), but up to this point only for
diagonal matrices $\mathsf \Xi$ in the distribution \eq{distribution}.

To obtain the same result for non-diagonal matrices $\mathsf \Xi$ in
\Eq{distribution}, we apply a transformation $\mathsf S$ (an
orthogonal $\mathcal N\times\mathcal N$ matrix with elements $S_{ij}$)
such that $\mathsf \Xi' =\mathsf S \cdot \mathsf \Xi \cdot \mathsf
S^T$ is diagonal. In this transformation, $r_i'=\sum_{j=1}^\mathcal N
S_{ij} r_j$, $v_i'=\sum_{j=1}^\mathcal N S_{ij} v_j$ and also $\Delta
r_i'= \sum_{j=1}^\mathcal N S_{ij} \Delta r_j$. Since $\mathsf \Xi'$
is diagonal, the cumulants of the primed displacements $\Delta r_i'$
satisfy the Theorem. The cumulants of the original displacements
$\Delta r_i=\sum_{j=1}^\mathcal N S_{ji}\Delta r'_j$ can be expressed
in terms of the primed ones using the multinomial expansion
\eq{starstar}, with the result
\begin{eqnarray}
  \kappa_{\{n_i\}}&=&
\mathop{\sum_{\{n_{1j}\}}}_{\sum_{j} n_{1j}=n_1}
\frac{n_1!}{n_{11}!\cdots n_{1\mathcal N}!}
\cdots
\mathop{\sum_{\{n_{\mathcal Nj}\}}}_{\sum_{j} n_{\mathcal Nj}=n_\mathcal N}
\frac{n_\mathcal N!}{n_{\mathcal N1}!\cdots n_{\mathcal N\mathcal N}!}
\nonumber
\\&&\times
\langle\!\langle
(S_{11}\Delta r_1')^{[n_{11}]};
\ldots;
(S_{\mathcal N1}\Delta r_\mathcal N')^{[n_{1\mathcal N}]};
\nonumber\\&&\quad\:\:
(S_{12}\Delta r_1')^{[n_{21}]};
\ldots;
(S_{\mathcal N2}\Delta r_\mathcal N')^{[n_{2\mathcal N}]};
\ldots;
(S_{\mathcal N\mathcal N}\Delta r_\mathcal N')^{[n_{\mathcal N\mathcal N}]}
\rangle\!\rangle
\nonumber\\
\end{eqnarray}
Since in each term on the right hand side the $\sum_{i,j=1}^\mathcal N
n_{ij} = \sum_{i=1}^\mathcal N n_{i}=n$ (cf.~\Eq{ndef}), and each term
contains the cumulants of primed displacements, which were already
shown to satisfy the Theorem, each term scales as $t^{2n}$ if $n>2$
and as $t^n$ if $n\leq2$, and therefore so does the sum. This proves
that \Eq{Generalized Theorem} of the Theorem holds for arbitrary
(positive symmetric) matrices~$\mathsf \Xi$.  \qed

\subsection{Expression for the coefficients in the Theorem}
\label{coefficients}

We will now show how one can determine the coefficients of the
$t^{2n}$ term of $\kappa_{\{n_i\}}$, i.e. $c_{\{n_i\}}$ in \Eq{Generalized
Theorem}.

For $n=1$ and $n=2$, it is straightforward to show that\cite{VanKampen}
\begin{subequations}
\label{cn}
\begin{eqnarray}
  c_{\{1,0,0,0,\ldots\}} &=& \average{v_1} \:=\: u_1\\
  c_{\{2,0,0,0,\ldots\}} &=& \average{V_1^2} \:=\: [\mathsf\Xi^{-1}]_{11}\\
  c_{\{1,1,0,0,\ldots\}} &=& \average{V_1V_2} \:=\: [\mathsf\Xi^{-1}]_{12}.
\end{eqnarray}
(and similarly for $c_{\{0,1,0,0,0,\ldots\}}$ and $c_{\{0,2,0,0,0,\ldots\}}$,
$c_{\{1,0,1,0,0,\ldots\}}$, $c_{\{0,1,1,0,0,\ldots\}}$, etc.).

To find $c_{\{n_i\}}$ for $n>2$ one can use a similar calculation of
$\kappa_{\{n_i\}}$ as used above, but taking one additional term
$X_{in+1}t^{n+1}$ in the time expansion of $\Delta r_i$ in
\Eq{lemmaA3} into account. Performing then the same kind of
manipulations as in the proof above, one arrives for $n>2$ at
\begin{multline}
c_{\{n_i\}} =
\mathop{
\mathop{
\sum_{\{n_{i\gamma}\geq0\}}
}_{\sum_{\gamma=1}^{n+1}n_{i\gamma}=n_i}
}_{\sum_{i=1}^\mathcal N\sum_{\gamma=1}^{n+1}\gamma n_{i\gamma}=2n} 
\left[\frac{n_i!}{\prod_{\gamma=1}^{n+1} n_{i\gamma}!}\right]
\\
\times
\cumulant{X_{11}^{[n_{11}]};\ldots;X_{1,n+1}^{[n_{1,n+1}]};
X_{21}^{[n_{21}]};\ldots;X_{\mathcal N,n+1}^{[n_{\mathcal N,n+1}]}}
.
\label{generalcoef}
\end{multline}
\end{subequations}
where $n$ was  defined in Eq. \eq{ndef}.

Note that for the case of a cumulant of one displacement, e.g.,
$\cumulant{\Delta r_i^n}$ in \Eq{Theorem0}, one has
\begin{equation}
c_n = c_{\{\ldots,0,n,0,\ldots\}}
\end{equation}
where the $n$ on the rhs\ is at the $i$th position.

Using \Eq{generalcoef}, we can give some examples of $c_{n}$ for $n=3$
and $n=4$:
\begin{eqnarray}
c_3 &=& 3\cumulant{V_i^{[2]};X_{i4}}+6\cumulant{V_i; X_{i2}; X_{i3}}
+\cumulant{X_{i2}^{[3]}}
\\
c_4 &=& 4\cumulant{V_i^{[3]};X_{i5}} +
6\cumulant{V_i^{[2]};X_{i3}^{[2]}}+12\cumulant{V_i^{[2]};X_{i2};X_{i4}}
\nonumber\\
&&+12\cumulant{V_i;X_{i2}^{[2]};X_{i3}} +\cumulant{X_{i2}^{[4]}}.
\end{eqnarray}
while an example of the coefficients $c_{\{n_i\}}$ for cumulants
involving different degrees of freedom, e.g.\  $\cumulant{\Delta
r_1;\Delta r_2;\Delta r_3}$ is, from \Eq{generalcoef},
\begin{align}
  c_{\{1,1,1,0,0\ldots\}}\: =\: &
\cumulant{V_1;V_2;X_{34}}
+
\cumulant{V_1;X_{22};X_{33}}
+
\cumulant{V_1;X_{23};X_{32}}
\nonumber\\&
+
\cumulant{V_1;X_{24};X_{31}}
+
\cumulant{X_{12};V_2;X_{33}}
+
\cumulant{X_{12};X_{22};X_{32}}
\nonumber\\&
+
\cumulant{X_{12};X_{23};V_3}
+
\cumulant{X_{13};V_2;X_{32}}
\nonumber\\&
+
\cumulant{X_{13};X_{22};V_3}
+
\cumulant{X_{14};V_2;V_3}.
\end{align}

We note that although the $X_{i\gamma}$ are useful to derive these
expressions for $c_{\{n_i\}}$, to evaluate them in practice requires
additional work. One would first need to write the $X_{i\gamma}$ as
$\tilde X_{i\gamma}(r^\mathcal N,V^\mathcal N,t=0)$ using \Eq{226p},
and work out the recursion relation \eq{newerrecursion}. Furthermore,
the cumulants would have to be written in terms of averages using
\Eq{cumintermsofmom}. The values of these averages will depend on the
system, i.e., the accelerations $a_i$, and their evaluation will in
general require a numerical approach, which we will explore in future
work.

\section*{Conclusions}
\label{conclusions}

In this paper, we have proved a mathematical theorem on the
correlations of the initial time displacements of particles (or in
general of coordinates associated with degrees of freedom) for a class
of dynamical systems whose main restriction is that the initial
distribution of the velocity-variables is a (multivariate) Gaussian,
independent of the position-variables.

Among the physical applications of this Theorem is the result that a
well-known short-time expansion of the Van Hove self-correlation is
well-behaved: each subsequent term in the expansion is smaller than
the previous one for small enough $t$, something which had been
suspected but not established before. It has also been shown on the
basis of the Theorem that in the studies of undercooled liquids and
glasses, using cumulants instead of the usual non-Gaussian parameters
may give more physical information.  Furthermore, it was shown that
the expansion used in the Green's functions theory is also
well-behaved if the velocity distributions are initially Gaussian, so
that this theory, which can describe non-equilibrium mass transport
processes on short time and length scales, has now been given a firmer
basis.

\section*{Acknowledgments}

The authors would like to thank Prof.~H.~van Beijeren and
Prof.~F.~Bonetto for carefully reading the manuscript and for their
helpful suggestions.  This work was supported by the Office of Basic
Energy Sciences of the US Department of Energy under grant number
DE-FG-02-88-ER13847.

\appendix

\section*{Appendix}
\renewcommand{\theequation}{A.\arabic{equation}}
\setcounter{equation}{0}

In the third step of the proof of the main Theorem, we needed the
following auxiliary theorem to prove the main Theorem.
\begin{theorem}
\label{cumulant with v power}
Let $V_i$ be a set of $\mathcal N$ statistically independent, zero-mean Gaussian
variables, collectively denoted by $V^\mathcal N$, and let $P_q$
($q=1\ldots Q$) be a set of $Q$ polynomials in $V^\mathcal N$ of total
degree $d_q$. Let $p_i$ ($i=1\ldots\mathcal N$) and $n_q$
($q=1\ldots Q$) be nonnegative integer numbers. Then if
\begin{equation}
\sum_{i=1}^\mathcal N p_i > \sum_{q=1}^Q n_q d_q
\end{equation}
and at least one $n_{q}\neq 0$, the
following cumulant vanishes:
\begin{equation}
  \thisbcumulant = 0.
\label{vanishing}
\end{equation} 
while if all $n_{q}$ are zero, one has
\begin{equation}
\cumulant{V_1^{[p_1]};\ldots; V_\mathcal N^{[p_\mathcal N]}}
=
\sum_{i=1}^\mathcal N 
\cumulant{V_i^{[2]}}\delta_{p_i2}\prod_{j\neq i}^\mathcal N
\delta_{p_j0}.
\label{nonvanishing}
\end{equation} 
\end{theorem}

\noindent
\emph{Proof of \Theorem{cumulant with v power}.} Since we have not
found a proof of this theorem in the literature, we will give it here,
but before we can prove \Theorem{cumulant with v power}, we need four
lemmas and the definition of a $\theta^\mathcal N$-modified average. This
definition will serve, in conjunction with \Lemma{theta and normal
cumulants} below, to construct a convenient generating function (which
takes the form of a $\theta^\mathcal N$-modified cumulant) for the
quantities $\thiscumulant$ that occur in \Theorem{cumulant with v
power}.  Working with this generating function will be more convenient
than trying to calculate each $\thiscumulant$ individually.  After
this generating function has been introduced, \Lemma{average with v
power} and \Lemma{theta average polynomial} are presented and proved,
which use the Gaussian nature of the $V_i$ to give properties of
averages and $\theta^\mathcal N$-modified averages of powers of the $V_i$,
which allow one to prove a polynomial property (\Lemma{theta cumulant
polynomial}) of the generating function of the quantities
$\thiscumulant$. This polynomial property will be the central
ingredient to complete the proof.

First we remark that below, $V^\mathcal N$ will always denote the same set
of $\mathcal N$ statistically independent, zero-mean Gaussian
distributed variables $V_i$ of \Theorem{cumulant with v power}.

The definition of $\theta^\mathcal N$-modified averages and cumulants is:

\begin{definition}
The $\theta^\mathcal N$-modified average of a general variable
$P$ (which below will always be a polynomial in $V^\mathcal N$) is defined as
\begin{equation}
  \average{P}_{\theta^\mathcal N} \equiv
\frac{\average{P\exp\sum_{i=1}^\mathcal N\theta_i V_i}}
{\average{\exp\sum_{i=1}^\mathcal N\theta_i V_i}}
=
e^{-\frac12\sum_{i=1}^\mathcal N\theta_i^2\langle{V_i^2}\rangle}
\baverage{P\exp\sum_{i=1}^\mathcal N\theta_i V_i}
\label{modavdef}
\end{equation}
where the $\theta_i$ ($i=1\ldots\mathcal N$) are real numbers.

Similarly, $\theta^\mathcal N$-modified moments are generally defined as
the $\theta^\mathcal N$-modified averages of powers of general variables
(i.e., functions of $V^\mathcal N$, and in the main text also of
$r^\mathcal N$), and $\theta^\mathcal N$-modified cumulants are defined as
having the same relation to $\theta^\mathcal N$-modified moments as normal
cumulants have to normal moments. $\theta^\mathcal N$-modified cumulants
are therefore also given through the $\theta^\mathcal N$-modified cumulant
generating function as:
\begin{equation}
\cumulant{P_1^{[n_1]};\ldots;P_Q^{[n_Q]}}_{\theta^\mathcal N}
\equiv
\prod_{q=1}^Q \dd[n_q]{\ }{(\mathrm{i} k_q)}
\log\baverage{\exp\sum_{q=1}^Q \mathrm{i} k_q
    P_q}_{\!\!\theta^\mathcal N}\bigg|_{\{k_q\}=0}.
\label{modcumdef}
\end{equation}
\end{definition}
\noindent
The generating function nature of the $\theta^\mathcal N$-modified cumulants 
follows from:

\begin{lemma}
{\emph{(relation between cumulants and $\theta^\mathcal N$-modified
cumulants)$\,$}}\footnote{For $\mathcal N=Q=n_1=1$ this lemma coincides with
the last exercise of section XVI.3 in Van Kampen's book\cite{VanKampen}}.
\label{theta and normal cumulants}
The $\theta^\mathcal N$-modified cumulants of a set of variables $P_1$, \ldots
$P_Q$ are related to the normal cumulants by
\begin{multline}
\cumulant{P_1^{[n_1]};\ldots;P_Q^{[n_Q]}}_{\theta^\mathcal N}
=
\sum_{p_1=0}^\infty\cdots\sum_{p_\mathcal N=0}^\infty 
\left[\prod_{i=1}^\mathcal N \frac{\theta_i^{p_i}}{p_i!}\right]
\\\times
\thiscumulant
\label{modcum}
\end{multline}
if at least one $n_q\neq0$, while it is zero otherwise.
\end{lemma}

\begin{proof}
For case in which at least one $n_q$ is nonzero, we start with the rhs\ of
\Eq{modcum} and use the expression for the cumulants in
\Eq{defcumvar}:
\begin{multline}
\sum_{p_1=0}^\infty\cdots\sum_{p_\mathcal N=0}^\infty  
\left[\prod_{i=1}^\mathcal N \frac{\theta_i^{p_i}}{p_i!}\right]
\thiscumulant
\\
=
\sum_{p_1=0}^\infty\cdots\sum_{p_\mathcal N=0}^\infty 
\prod_{i=1}^\mathcal N 
\frac{\theta_i^{p_i}}{p_i!}
\dd[p_i]{\ }{(\mathrm{i} k_i')}
\prod_{q=1}^Q \dd[n_q]{\ }{(\mathrm{i} k_q)}
\\
\times
\log\baverage{\exp\Big[\sum_{i=1}^\mathcal N \mathrm{i} k'_i V_i+\sum_{q=1}^Q
\mathrm{i} k_q  P_q\Big]}
\Bigg|_{\{k_q\}=\{k'_i\}=0}
\end{multline}
We recognize the Taylor series, i.e., that
$\sum_{p_i=0}^{\infty}([-\mathrm i\theta_i]^{p_i}/p_i!)\, \partial^{p_i} f(k'_i)/\partial
{k_i'^n}\Big|_{k_i'=0}=f(-\mathrm i\theta_i)$ to write this as
\begin{multline}
\sum_{p_1=0}^\infty\cdots\sum_{p_\mathcal N=0}^\infty 
\left[\prod_{i=1}^\mathcal N \frac{\theta_i^{p_i}}{p_i!}\right]
\thiscumulant
\\
=
\prod_{q=1}^Q \dd[n_q]{\ }{(\mathrm{i} k_q)}
\log\baverage{\exp\Big[\sum_{i=1}^\mathcal N \theta_i V_i+\sum_{q=1}^Q
\mathrm{i} k_q  P_q\Big]}
\Bigg|_{\{k_q\}=0}
\end{multline}
Using definition \eq{modavdef}, we obtain
\begin{multline}
\sum_{p_1=0}^\infty\cdots\sum_{p_\mathcal N=0}^\infty 
\left[\prod_{i=1}^\mathcal N \frac{\theta_i^{p_i}}{p_i!}\right]
\thiscumulant
\\
=
\prod_{q=1}^Q \dd[n_q]{\ }{(\mathrm{i} k_q)}
\bigg[
\log\baverage{\exp\sum_{q=1}^Q
\mathrm{i} k_q  P_q}_{\theta^\mathcal N}
+ \frac12\sum_{i=1}^\mathcal N\theta_i^2\average{V_i^2}\bigg]
\Bigg|_{\{k_q\}=0}
\label{thistoo}
\end{multline}
Since we are considering the case that at least one $n_q$ is
nonzero, the contribution from the $k_q$-independent
$\frac12\sum_{i=1}^\mathcal N\theta_i^2\average{V_i^2}$ vanishes when
$\partial/\partial({\mathrm i k_q})$ acts on it. The remainder on the
rhs\ of \Eq{thistoo} is
by definition \eq{modcumdef} equal to the lhs\ of \Eq{modcum}.

The case $n_1=\cdots=n_Q=0$ is trivial, for then $\cumulant{P_1^{n_1}
\ldots P_Q^{n_Q}}\big._{\theta^\mathcal N} =$ since a zeroth cumulant
is always zero.

\end{proof}

A consequence of this lemma, i.e., of \Eq{modcum}, is that repeated
derivatives with respect to $\theta_i$ of
$\cumulant{P_1^{[n_1]};\ldots; P_Q^{[n_Q]}}_{\theta^\mathcal N}$ taken
at $\theta_i=0$ generate the $\thiscumulant$.\vspace{2mm}

\ \\
The second lemma concerns the average of a product of powers of the $V_i$
and a single polynomial in $V^\mathcal N$.

\begin{lemma}{\emph{(Averages of powers of $V_i$ times a
      polynomial in $V^\mathcal N$)}}
\label{average with v power}
Let $P$ be a polynomial function of total degree $d$ in $V^\mathcal N$.
Given a set of $\mathcal N$ nonnegative integers $s_i$, collectively
denoted by $s^\mathcal N$, and a set of $\mathcal N$ ``parities'' $\delta_i$
with each $\delta_i$ zero or one, one can write
\begin{equation}
\baverage{\prod_{i=1}^\mathcal N V_i^{2s_i+\delta_i} P} 
=
 P_{s^*}(\{s_j\}) \prod_{i=1}^\mathcal N  (2s_i+2\delta_i-1)!!
\average{V_i^2}^{s_i}
\label{odd}
\end{equation}
where $P_{s^*}(\{s_j\})$ is a polynomial in the $s_j$ of total
degree $s^*$, which satisfies
\begin{equation}
s^* \leq \frac{d-\sum_{i=1}^\mathcal N\delta_i}{2}.
\label{sstar}
\end{equation}
\end{lemma}

\begin{proof}
We start by writing out the polynomial $P$ as
\begin{equation}
  P =  \mathop{\sum_{\{p_i\}}}_{\sum_{i=1}^\mathcal N p_i\leq d}
 b_{\{p_i\}} \prod_{i=1}^\mathcal N V_i^{p_i}.
\end{equation}
Of these terms, only those with the same ``parity'' as the $\delta_i$
(i.e., if $\delta_i=0$, $p_i$ is even and if $\delta_i=1$,
$p_i$ is odd) contribute to the average $\average{\prod_{i=1}^\mathcal N
V_i^{2s_i+\delta_i} P}$, due to the even nature of the distribution of
the $V_i$. Thus, one can write $p_i=2s'_i+\delta_i$ and obtain
\begin{equation}
\baverage{\prod_{i=1}^\mathcal N V_i^{2s_i+\delta_i} P}
=  \mathop{\sum_{\{s'_i\}}}_{\sum_{i=1}^\mathcal N
  (2s'_i+\delta_i)\leq d}
 b_{\{2s'_i+\delta_i\}} \prod_{i=1}^\mathcal N \average{V_i^{2(s_i+s_i'+\delta_i)}}.
\end{equation}
Using $\average{V^{2s}_i}=(2s-1)!!\average{V_i^2}^s$\cite{Cramer46},
we obtain
\begin{align}
\baverage{\prod_{i=1}^\mathcal N V_i^{2s_i+\delta_i} P}
&=  \mathop{\sum_{\{s'_i\}}}_{\sum_{i=1}^\mathcal N
  (2s'_i+\delta_i)\leq d}
 b_{\{2s'_i+\delta_i\}} \prod_{i=1}^\mathcal N (2s_i+2s_i'+2\delta_i-1)!!
\average{V_i^2}^{s_i+s'_i+\delta_i}
\nonumber\\
&=
 \mathop{\sum_{\{s'_i\}}}_{\sum_{i=1}^\mathcal N
  (2s'_i+\delta_i)\leq d}
 b_{\{2s'_i+\delta_i\}} \prod_{i=1}^\mathcal N
 \frac{(2s_i+2s_i'+2\delta_i-1)!!}
  {(2s_i+2\delta_i-1)!!}
\average{V_i^2}^{s_i'+\delta_i}
\nonumber\\
&\qquad\qquad\times
\prod_{i=1}^\mathcal N
  (2s_i+2\delta_i-1)!!\average{V_i^2}^{s_i}
\end{align}
This is of the form stated in \Eq{odd}, where
\begin{equation}
 P_{s^*}(\{s_j\}) = 
 \mathop{\sum_{\{s'_i\}}}_{\sum_{i=1}^\mathcal N
  (2s'_i+\delta_i)\leq d}
 b_{\{2s'_i+\delta_i\}} \prod_{i=1}^\mathcal N
\prod_{s''_i=1}^{s_i'}
(2s_i+2s_i''+2\delta_i-1)
\average{V_i^2}^{s_i'+\delta_i}
\label{ook}
\end{equation}
This shows that $P_{s^*}$ is a polynomial in $s^\mathcal N$, since it
depends on the $s_i$ only through finitely many factors
$(2s_i+2s_i''+2\delta_i-1)$. In fact, for each term in \Eq{ook}, the
number of such factors is $\sum_{i=1}^\mathcal N s'_i$. Because of the
restriction on the sum over $\{s_i'\}$ in \Eq{ook}, this number
$\sum_{i=1}^\mathcal N s'_i$ is less than or equal to $(d-\sum_{i=1}^\mathcal N
\delta_i)/2$ for each term. The total degree $s^*$ of the polynomial
$P_{s^*}(\{s_j\})$, which is the maximum of this number over all the
terms, therefore also satisfies $s^*\leq (d-\sum_{i=1}^\mathcal N
\delta_i)/2$.
\end{proof}

The third lemma concerns a polynomial property of the
$\theta^\mathcal N$-modified averages of polynomials $P$ in $V^\mathcal N$.

\begin{lemma}{\emph{(polynomial property of the $\theta^\mathcal N$-modified average of a polynomial
 in $V^\mathcal N$)}}
\label{theta average polynomial}
Let $P$ be a polynomial of total degree $d$ in $V^\mathcal N$.  Then the
$\theta^\mathcal N$-modified average of $P$ is a polynomial in
$\theta^\mathcal N$ of
at most the same total degree $d$.
\end{lemma}

\begin{proof}
a) Consider first the case that $P$ is a polynomial in the $V_i$ of
total degree $d$ and  of
definite ``parity'' for each $i$, i.e., that it either changes
sign or remains unchanged when $V_i$ is replaced by $-V_i$. Define the
numbers $\delta_i$ such that $\delta_i=0$ if $P$ is even in $V_i$ and
$\delta_i=1$ if it is odd. By the definition of the
$\theta^\mathcal N$-modified average in \Eq{modavdef}, expanding the
exponent, and using that only even powers of $V_i$ have a non-zero
average, one gets
\begin{eqnarray}
  \average{P}_{\theta^\mathcal N} &=&
e^{-\frac12\sum_{i=1}^\mathcal N\theta_i^2\langle{V_i^2}\rangle}
\nonumber\\&&\times
\sum_{s_1=0}^\infty\cdots\sum_{s_\mathcal N=0}^\infty 
\left[\prod_{i=1}^\mathcal N
\frac{\theta_i^{2s_i+\delta_i}}
{(2s_i+\delta_i)!}
\right]
\baverage{\prod_{i=1}^\mathcal N V_i^{2s_i+\delta_i} P}
\end{eqnarray}
Using \Eq{odd} of \Lemma{average with v power} and that
$(2s_i-1+2\delta_i)!!/(2s_i+\delta_i)!= 1/(2^{s_i} s_i!)$, we obtain
\begin{multline}
  \average{P}_{\theta^\mathcal N} =
 e^{-\frac12\sum_{i=1}^\mathcal N\theta_i^2\langle{V_i^2}\rangle}
 \sum_{\{s_i\geq 0\}} 
 \prod_{i=1}^\mathcal N \frac{\theta_i^{2s_i+\delta_i}
 \average{V_i^2}^{s_i+\delta_i}}{2^{s_i}s_i!}
 P_{s^*}(\{s_j\})
\\
 =
 \left[\prod_{i=1}^\mathcal N[\average{V_i^2}\theta_i]^{\delta_i}\right]
 e^{-\frac12\sum_{i=1}^\mathcal N\theta_i^2\langle{V_i^2}\rangle}
 \sum_{\{s_i\geq 0\}} 
 P_{s^*}(\{s_j\})
 \prod_{i=1}^\mathcal N 
 \frac{[\frac12\theta_i^2\average{V_i^2}]^{s_i}}
 {s_i!}
\label{A18}
\end{multline}
Note that the polynomial $P_{s^*}$ in $s_j$ inside the $\{s_i\}$
summation can be generated from an expression that does not have this
polynomial by applying operators $\theta_j^2
(\partial/\partial\theta_j^2)$, e.g.,
\begin{eqnarray}
 \sum_{\{s_i\geq 0\}} s_j \prod_{i=1}^\mathcal N
 \frac{[\frac12\theta_i^2\average{V_i^2}]^{s_i}}{s_i!}
 &=&
 \theta_j^2
 \dd{}{{\theta_j^2}}\sum_{\{s_i\geq 0\}} \prod_{i=1}^\mathcal N
 \frac{[\frac12\theta_i^2\average{V_i^2}]^{s_i}}{s_i!} ,
 \nonumber\\
 &=& \theta_j^2
 \dd{}{{\theta_j^2}}
\: e^{\frac12\sum_{i=1}^\mathcal N\theta_i^2\langle{V_i^2}\rangle}.
\end{eqnarray}
and in general
\begin{equation}
\sum_{\{s_i\geq 0\}} 
 P_{s^*}(\{s_j\})
 \prod_{i=1}^\mathcal N 
 \frac{[\frac12\theta_i^2\average{V_i^2}]^{s_i}}
 {s_i!}
 =
 P_{s^*}\Big(\{\theta^2_j \ddh{}{{\theta_j^2}}\}\Big)
 e^{\frac12\sum_{i=1}^\mathcal N\theta_i^2\langle{V_i^2}\rangle}.
\end{equation}
Combining this with \Eq{A18} gives
\begin{equation}
  \average{P}_{\theta^\mathcal N} =
\left[\prod_{i=1}^\mathcal N[\average{V_i^2}\theta_i]^{\delta_i}\right]
e^{-\frac12\sum_{i=1}^\mathcal N\theta_i^2\langle{V_i^2}\rangle}
P_{s^*}\Big(\{\theta^2_j \dd{}{{\theta_j^2}}\}\Big)
e^{\frac12\sum_{i=1}^\mathcal N\theta_i^2\langle{V_i^2}\rangle}.
\label{A20}
\end{equation}
In this expression, the differential operators
$(\partial/\partial\theta_j^2)$ ``bring down'' factors of $\theta_j^2$
from the exponent to its right. That exponent itself is then canceled
by the exponent to the left of the $P_{s^*}$ operator, so that only a polynomial in
$\theta^\mathcal N$ is left. The total degree of this polynomial is
twice the maximum number of factors brought down by the operators
(twice because the squares of the $\theta_j$ are brought down). This
maximum number is simply the total degree of $P_{s^*}$, i.e.,
$s^*$. Counting finally also the powers of $\theta_i$ of the first
product in \Eq{A20}, we see that $\average{P}_{\theta^\mathcal N}$ is a
polynomial in $\theta^\mathcal N$ of total degree
$d'=2s^*+\sum_{i=1}^\mathcal N \delta_i$.  But \Lemma{average with v
power}, in particular \Eq{sstar}, says that $s^*\leq
(d-\sum_{i=1}^\mathcal N\delta_i)/2$, so that for a polynomial $P$ of
definite parity, its total degree $d'$ satisfies $d'\leq d$.

b) To show that the same is true for a general polynomial, note that
any polynomial $P$ can always be written as a sum of polynomials of
definite parity. Since the $\theta^\mathcal N$-modified average of that sum
is the sum of the $\theta^\mathcal N$-modified average of each term, and
each term is a polynomial in $\theta^\mathcal N$ of total degree
$d'\leq d$, $\average{P}_{\theta^\mathcal N}$ is also a polynomial in
$\theta^\mathcal N$ of total degree $d'\leq d$.
\end{proof}

The next and final lemma we need before we can prove \Theorem{cumulant
with v power} concerns $\theta^\mathcal N$-modified cumulants of several
polynomials in $V^\mathcal N$.

\begin{lemma}
{\emph{($\theta^\mathcal N$-modified cumulants of polynomials in $V^\mathcal N$)}}
\label{theta cumulant polynomial}
Let $P_q$ ($q=1\ldots Q$) be a set of polynomials of total degree
$d_q$ in the $V^\mathcal N$. Then the $\theta^\mathcal N$-modified
cumulant $\cumulant{P_1^{[n_1]} ; \ldots ;
P_Q^{[n_Q]}}_{\theta^\mathcal N}$ is a polynomial in $\theta^\mathcal
N$ of a total degree of at most $d =\sum_{q=1}^{Q}n_q d_q$.
\end{lemma}

\begin{proof}
Because the $\theta^\mathcal N$-modified cumulants are defined
formally in precisely the same way as normal cumulants [i.e.,
\Eq{modcumdef} vs.\ \Eq{defcumvar}], the relation \Eq{cumintermsofmom}
between cumulants and moments applies to the $\theta^\mathcal
N$-modified cumulants and moments as well. According to that relation,
the $\theta^\mathcal N$-modified cumulant
$\cumulant{P_1^{[n_1]};\ldots; P_Q^{[n_Q]}}_{\theta^\mathcal N}$
can be expressed as a sum of terms each of which contains a product of
the moments $\average{P_1^{\ell_{1}}\cdots
P_Q^{\ell_{Q}}}_{\theta^\mathcal N}$ raised to the power
$p_{\{\ell\}}$ (where the product is over all possible sets
$\{\ell\}=\{l_1,\ldots, l_Q\}$ with the restrictions stated in
\Eq{cumintermsofmom}). Because $P_q$ is required by the conditions of
the lemma to be a polynomial in $V^\mathcal N$ of total degree $d_q$,
$P_q^{\ell_q}$ is a polynomial in $V^\mathcal N$ of total degree
$\ell_qd_q$, and the product $P_1^{\ell_{1}} \cdots P_Q^{\ell_Q}$ is a
polynomial in $V^\mathcal N$ of total degree $\sum_{q=1}^Q \ell_q
d_q$.  According to \Lemma{theta average polynomial}, each
$\theta^\mathcal N$-modified moment $\average{P_1^{\ell_{1}} \cdots
P_Q^{\ell_Q}}_{\theta^\mathcal N}$ is then a polynomial in
$\theta^\mathcal N$ of a total degree of at most $\sum_{q=1}^Q \ell_q
d_q$. Its $p_{\{\ell\}}$-th power in \Eq{cumintermsofmom} is then a
polynomial in $\theta^\mathcal N$ of total degree
$p_{\{\ell\}}\sum_{q=1}^Q \ell_q d_q$ (at most), and the product (over
$\{\ell\}$) that occurs on the rhs\ of \Eq{cumintermsofmom} is of
total degree $d\equiv\sum_{\{\ell\}}p_{\{\ell\}}\sum_{q=1}^Q \ell_q
d_q$ (at most).  Since $p_{\{\ell\}}$ is summed over in
\Eq{cumintermsofmom} with the restriction that
$\sum_{\{\ell\}}p_{\{\ell\}}\ell_q=n_q$, the total degree $d$ of the
expressions $\prod_{\{\ell\}} \average{P_1^{\ell_{1}}\cdots
P_Q^{\ell_Q}}_{\theta^\mathcal N}^{p_{\{\ell\}}}$ can be rewritten as
$d=\sum_{q=1}^Q n_q d_q$ (at most).  Each term in the sum over
$p_{\{\ell\}}$ on the rhs\ of \Eq{cumintermsofmom} is therefore (at
most) of this total degree $d$, so that the full expression, i.e.
$\cumulant{P_1^{[n_1]};\ldots; P_Q^{[n_Q]}}_{\theta^\mathcal N}$,
is also a polynomial in $\theta^\mathcal N$ of a total degree of at
most~$d$.
\end{proof}

\subsubsection*{\rm\em Conclusion of the proof of \Theorem{cumulant with v power}}

We can now finish the proof of \Theorem{cumulant with v power}.  To
consider $\langle\!\langle V_1^{[p_1]};\ldots;V_{\mathcal
N}^{[p_{\mathcal N}]} ;$ $P_1^{[n_1]};\ldots;
P_Q^{[n_Q]}\rangle\!\rangle$ as on the lhs\ of \Eq{vanishing} in the
case that at least one $n_q\neq 0$, we look at its generating
function, i.e., $\cumulant{P_1^{[n_1]};\ldots;
P_Q^{[n_Q]}}_{\theta^\mathcal N}$.  According to \Eq{modcum} of
\Lemma{theta and normal cumulants}, this generating function admits a
power series in $\theta^\mathcal N$. At the same time, according to
\Lemma{theta cumulant polynomial} (whose proof required Lemmas
\ref{average with v power} and \ref{theta average polynomial}),
$\cumulant{P_1^{[n_1]};\ldots; P_Q^{[n_Q]}}_{\theta^\mathcal N}$
is given by a polynomial in $\theta^\mathcal N$ of a total degree of
at most $d =\sum_{q=1}^Q n_q d_q$, so its power series in
$\theta^\mathcal N$ in \Eq{modcum} terminates (at the latest) after
$\sum_{i=1}^\mathcal N p_i=d$. Since these lemmas should be valid for
all $\theta^\mathcal N$, each coefficient of the terms with
$\sum_{i=1}^\mathcal N p_i>d$ must be zero, i.e., using \Eq{modcum} of
\Lemma{theta and normal cumulants}, $\thiscumulant=0$,
which is \Eq{vanishing} of \Theorem{cumulant with v power}. Finally,
for the case $n_q=0$ for all $q=1\ldots Q$, \Eq{nonvanishing} of
\Theorem{cumulant with v power} follows simply from the zero-mean
Gaussian nature of the $V_i$ and their statistical independence.  \qed

\end{document}